\documentclass[12pt]{article}
\pdfoutput=1
\usepackage[a4paper]{geometry}
\usepackage{jheppub, amsmath,amssymb,amsfonts,amsxtra, mathrsfs, makeidx,graphics,graphicx,amsthm,epsfig, ytableau,bm,longtable,float, color,tikz,mathtools,xfrac,footnote,rotating, lscape, makecell, environ,mathtools, empheq, colortbl, xcolor,enumerate}

\usetikzlibrary{positioning}
\usetikzlibrary{chains}
\usetikzlibrary{arrows, arrows.meta ,fit,decorations.pathreplacing}
\tikzstyle{every picture}+=[remember picture]
\tikzstyle{na} = [baseline]

\usetikzlibrary{arrows, decorations.markings, calc, fadings, decorations.pathreplacing, patterns, decorations.pathmorphing, positioning}
\tikzset{>={Latex[width=1.5mm,length=1.5mm]}}

\usepackage{mdframed}

\def\node#1#2{\overset{#1}{\underset{#2}{{\color{gray} \bullet}}}}

\def\node#1#2{\overset{#1}{\underset{#2}{\circ}}}

\tikzstyle{every picture}+=[remember picture]
\tikzstyle{na} = [baseline=-.5ex]

\usepackage{bbm}
\usepackage{subfigure}

\newcommand{\eg}{\textit{e.g.}}

\newcommand{\ie}{\textit{i.e.}}

\numberwithin{equation}{section}
\newcommand{\bes}[1]{\begin{equation} \begin{split} #1\end{split} \end{equation}}

\newcommand{\nn}{\nonumber}

\newcommand{\be}{\begin{equation}} \newcommand{\ee}{\end{equation}}
\newcommand{\bea}{\begin{equation} \begin{aligned}} \newcommand{\eea}{\end{aligned} \end{equation}}

\def\tilde{\widetilde}

\def\bar{\overline}

\def\rt2{\sqrt{2}}
\def\half {{1 \over 2}}

\def\Tr{\mathop{\rm Tr}}
\def\tr{\mathop{\rm tr}}

\def\CI{{\cal I}}

\def\CN{{\cal N}}

\def\1{{\ds 1}}

\def\repa{\raise4pt\hbox{$\square$}\mkern-14mu\raise-4pt\hbox{$\square$}}
\def\repab{\overline{\raise4pt\hbox{$\square$}\mkern-14mu\raise-4pt\hbox{$\square$}\mkern-1mu}}

\def\smileface{\ensuremath{\hbox{\large$\bigcirc$}\mkern-15mu\raise-1pt\hbox{\scriptsize$\smallsmile$}%
\mkern-10mu\raise4pt\hbox{..}\mkern4mu}}
\def\frownface{\ensuremath{\hbox{\large$\bigcirc$}\mkern-15mu\raise-1pt\hbox{\scriptsize$\smallfrown$}%
\mkern-10mu\raise4pt\hbox{..}\mkern4mu}}

\newcommand{\ba}{\begin{array}}
\newcommand{\ea}{\end{array}}
\newcommand{\bi}{\begin{itemize}}
\newcommand{\ei}{\end{itemize}}
\def\vec#1{\bm{#1}}
\def\bea#1\eea{\allowdisplaybreaks \begin{align}#1\end{align}}
 \newcommand{\ben}{\begin{enumerate}}
\newcommand{\een}{\end{enumerate}}
\newcommand{\bean}{\begin{eqnarray*}}
\newcommand{\eean}{\end{eqnarray*}}
\newcommand{\eref}[1]{(\ref{#1})}

\newcommand{\PE}{\mathop{\rm PE}}

\newcommand{\comment}[1]{}

\definecolor{light-gray}{gray}{0.7}

\newcommand{\purple}{\color{brown}}
\newcommand{\blue}{\color{blue}}

\newcommand{\red}{\color{red}}

\def\aup#1 {\overset{#1}{\uparrow} \, \overset{\tilde{#1}}{\downarrow}}

\tikzset{snake it/.style={decorate, decoration={snake, amplitude=.4mm, segment length=2mm,
                       post length=0mm,pre length=0mm}}}
                       
\def\S{{\sf S}}

\def\S{{\mathbb S}}

\newcommand{\fug}{\tau}

\def\gd{\delta}

\def\gs{\sigma}

\def\Gp{\Phi}

\tikzset{->-/.style={decoration={
  markings,
  mark=at position #1 with {\arrow{>}}},postaction={decorate}}}    
\tikzset{-<-/.style={decoration={
  markings,
  mark=at position #1 with {\arrow{<}}},postaction={decorate}}}          
  
\title{Symmetry enhancement and duality walls in 5d gauge theories}
\author[a,b]{Ivan Garozzo,} 
\author[b,c]{Noppadol Mekareeya,}
\author[a,b]{Matteo Sacchi,}
\author[a,b]{and Gabi Zafrir}
\affiliation[a]{Dipartimento di Fisica, Universit\`a di Milano-Bicocca, \\ Piazza della Scienza 3, I-20126 Milano, Italy}
\affiliation[b]{INFN, sezione di Milano-Bicocca, \\Piazza della Scienza 3, I-20126 Milano, Italy}
\affiliation[c]{Department of Physics, Faculty of Science, \\
Chulalongkorn University, Phayathai Road, \\
Pathumwan, Bangkok 10330, Thailand}
\emailAdd{ivangarozzo@gmail.com}
\emailAdd{n.mekareeya@gmail.com}
\emailAdd{m.sacchi13@campus.unimib.it}
\emailAdd{gabi.zafrir@unimib.it}
\abstract{Gauge theories in four dimensions can exhibit interesting low energy phenomena, such as infrared enhancements of global symmetry.  We explore a class of 4d $\mathcal{N}=1$ gauge theories arising from a construction that is motivated by duality walls in 5d gauge theories. Their quiver descriptions bear a resemblance to 4d theories obtained by compactifying 6d $\mathcal{N}=(1,0)$ superconformal field theories on a torus with fluxes, but with lower number of flavours and different number of gauge singlets and superpotentials. One of the main features of these theories is that they exhibit a flavour symmetry enhancement, and with supersymmetry enhancement for certain models, in the infrared. Properties of the superconformal fixed points of such theories are investigated in detail.}
\begin{document}
\maketitle

\section{Introduction}
Enhancement of global symmetry in the infrared is one of the most fascinating phenomena in quantum field theory.  This can occur when certain operators become conserved currents at the fixed point in the infrared (IR), and make the global symmetry in the IR larger than that in the ultraviolet (UV).  One of the reasons that makes the symmetry enhancement intriguing is due to the lack of a general principle and mechanism to explain such a phenomenon, especially in four spacetime dimensions.  Nevertheless, supersymmetry allows one to study the enhancement of symmetry in a more tractable fashion.  This is due to the presence of quantities that do not depend on the renormalisation group flow \cite{Festuccia:2011ws}, such as the supersymmetric index in four dimensions \cite{Kinney:2005ej,Romelsberger:2005eg, Dolan:2008qi}, that enable us to easily extract information about the conserved currents at the strongly coupled fixed point by a calculation in the weakly coupled regime.

In this paper we focus on a class of 4d $\CN=1$ supersymmetric gauge theories arising from a construction that is motivated by duality walls in 5d $\CN=1$ gauge theories \cite{Gaiotto:2015una}.  Their quiver descriptions are very similar to those studied in \cite{Kim:2017toz, Kim:2018bpg, Kim:2018lfo, Zafrir:2018hkr, Pasquetti:2019hxf}, but with lower number of flavours and different number of gauge singlets and superpotentials.  One of the main features of such gauge theories is that they exhibit a flavour symmetry enhancement, as well as supersymmetry enhancement for some models, in the IR.  Those with supersymmetry enhancement can be regarded as the complements to the models considered in \cite{Maruyoshi:2016tqk, Maruyoshi:2016aim, Agarwal:2016pjo, Agarwal:2017roi, Benvenuti:2017bpg}\footnote{In fact, in section \ref{sec:A1D4}, we consider a theory that is Seiberg dual to the one explored in section 3.2 of \cite{Agarwal:2016pjo} and section 2.1 of \cite{Benvenuti:2017bpg}.}.  In the following, we describe the construction of the aforementioned 4d $\CN=1$ gauge theories in detail. 

\subsection*{Duality walls in 5d $\CN=1$ gauge theories}
Four dimensional theories associated with duality walls in 5d $\CN=1$ gauge theories were proposed and studied in \cite{Gaiotto:2015una}.  
For definiteness, let us consider 5d $\CN=1$ $SU(N)$ gauge theory with $2N$ flavours of fundamental hypermultiplets, and Chern-Simons level zero.  For $N=2$ this 5d theory has a UV completion as a 5d $\CN=1$ SCFT with an enhanced flavour symmetry $E_5 \cong SO(10)$ \cite{Seiberg:1996bd}, whereas for $N\geq 3$ the UV completion is a 5d $\CN=1$ SCFT with an enhanced symmetry $SU(2N) \times SU(2)^2$ \cite{Bergman:2013aca,Mitev:2015pty}.  The 4d $\CN=1$ theory in question is a Wess--Zumino model that can be represented by the following quiver diagram \cite[fig. 12]{Gaiotto:2015una}:
\be \label{basic5dwall}
\begin{tikzpicture}[baseline]
\tikzstyle{every node}=[font=\footnotesize]
\node[draw, rectangle] (node1) at (-2,-1) {$N$};
\node[draw, rectangle] (node2) at (2,-1) {$N$};
\node[draw, rectangle] (sqnode) at (0,1) {$2N$};
 \draw[transform canvas={yshift=0pt}, , ->-=0.5] (node1) to  node[below] {\blue $D$} node[below, xshift=-0.5cm] {$\red F$} node[xshift=-0.5cm] {$ \red \times$} (node2) ;
\draw[draw=black,solid, -<-=0.5]  (node1) to node[left] {\blue $L$} (sqnode);
\draw[draw=black,solid, -<-=0.5]  (sqnode) to node[right] {\blue $R$} (node2);
\end{tikzpicture}
\ee
Through out the paper, a white node labelled by $n$ denotes the group $SU(n)$.  We denote each factor of the gauge symmetry by a circular node and the flavour symmetry by a rectangular node.  The superpotential is taken to be
\be
W= L^i_a D^a_{a'} R^{a'}_i+ F \left( \epsilon_{a_1 \ldots a_N} \epsilon^{a'_1 \ldots a'_N}  D^{a_1}_{a'_1}  \cdots D^{a_N}_{a'_N} \right)  ~,
\ee
where the unprimed indices $a, a_1, a_2, \ldots = 1,\ldots, N$ are those of the left $SU(N)$ node; the primed indices $a', a_1', a_2', \ldots =1,\ldots, N$ are those of the right $SU(N)$ node; and the indices $i,j=1,\ldots, 2N$ are those of the top $SU(2N)$ node.   The duality wall imposes the Neumann boundary condition for the $SU(N)$ gauge theory on the two sides of the wall, and thus gives rise to the bottom left and bottom right $SU(N)$ nodes in \eref{basic5dwall}.  The top $SU(2N)$ node comes from the 5d flavour symmetry.  Using \eref{basic5dwall} as a building block, one can construct a number of interesting gauge theories by simply gluing the building blocks together.  For example, one can concatenate two duality walls in this 5d theory, and the corresponding 4d theory has the following quiver description \cite[fig. 13]{Gaiotto:2015una}:
\be
\scalebox{0.8}{
\begin{tikzpicture}[baseline]
\tikzstyle{every node}=[font=\footnotesize]
\node[draw, rectangle] (node1) at (-3,-1) {$N$};
\node[draw, circle] (c) at (0,-1) {$N$};
\node[draw, rectangle] (node2) at (3,-1) {$N$};
\node[draw, rectangle] (sqnode) at (0,1) {$2N$};
 \draw[transform canvas={yshift=0pt}, , ->-=0.5] (node1) to  node[below] {\blue $D_1$} node[below, xshift=-0.5cm] {$\red F_1$} node[xshift=-0.5cm] {$ \red \times$} (c) ;
  \draw[transform canvas={yshift=0pt}, , -<-=0.5] (c) to  node[below] {\blue $D_2$} node[below, xshift=-0.5cm] {$\red F_2$} node[xshift=-0.5cm] {$ \red \times$} (node2) ;
\draw[draw=black,solid, -<-=0.5]  (node1) to node[left] {\blue $L$} (sqnode);
\draw[draw=black,solid, ->-=0.5]  (c) to node[left] {\blue $V$} (sqnode);
\draw[draw=black,solid, ->-=0.5]  (sqnode) to node[right] {\blue $R$} (node2);
\node[draw=none] at (0,-2) {$W=LD_1 V+ RD_2V + F_1 D_1^N + F_2 D_2^N$};
\end{tikzpicture}}
\ee

\subsection*{The $E$-string theory on Riemann surfaces with fluxes}
Theory \eref{basic5dwall} can be modified in order to describe 4d theories associated with a duality wall in other 5d theories.  An interesting modification was studied in \cite{Kim:2017toz} in the context of the compactification of 6d rank-one $E$-string theory on Riemann surfaces with fluxes.  In that reference, the case of $N=2$ is investigated and the top $SU(2N)=SU(4)$ node is replace by $SU(8)$; see \cite[fig. 10(a)]{Kim:2017toz}:  
\be \label{basicEstring}
\begin{tikzpicture}[baseline]
\tikzstyle{every node}=[font=\footnotesize]
\node[draw, rectangle] (node1) at (-2,-1) {$2$};
\node[draw, rectangle] (node2) at (2,-1) {$2$};
\node[draw, rectangle] (sqnode) at (0,1) {$8$};
 \draw[transform canvas={yshift=0pt}] (node1) to  node[below] {\blue $D$} node[below, xshift=-0.5cm] {$\red F$} node[xshift=-0.5cm] {$ \red \times$} (node2) ;
\draw[draw=black,solid, -<-=0.5]  (node1) to node[left] {\blue $L$} (sqnode);
\draw[draw=black,solid, -<-=0.5]  (sqnode) to node[right] {\blue $R$} (node2);
\node[draw=none] at (0,-2) {$W=L DR+  F(DD)$};
\end{tikzpicture}
\ee
The corresponding 5d $\CN=1$ theory is the $SU(2)$ gauge theory with $8$ flavours, whose UV completion is the 6d rank-one $E$-string theory \cite{Ganor:1996mu, Seiberg:1996vs, Ganor:1996pc}.  The flavour symmetry of theory \eref{basicEstring} is $SU(2)^2 \times SU(8) \times U(1)_F \times U(1)$, where $SU(8) \times U(1)_F$ is a subgroup of the $E_8$ symmetry of the $E$-string theory.
Theory \eref{basicEstring} can be interpreted as coming from the compactification of the rank-one $E$-string theory on a two punctured sphere (\ie~ a tube) with a particular choice of 6d flux that breaks the $E_8$ symmetry to $E_7  \times U(1)_F$.  Note that each puncture brings about an $SU(2)$ symmetry and breaks $E_7 \times U(1)_F$ to $SU(8) \times U(1)_F$.   From the 5d perspective, the $U(1)_F$ symmetry implies the presence of a {\it duality domain wall} such that the mass parameter for $U(1)_F$ flips its sign as we go from one side of the wall to the other.  As discussed in \cite[sec. 3]{Kim:2017toz}, one way to see the $E_7 \times U(1)_F$ symmetry is to glue the two punctures together (\ie~ close the tube) to form a torus.  The corresponding 4d theory can be obtained by taking two copies of \eref{basicEstring} and `self-gluing' by identifying their $SU(8)$ nodes and commonly gauging each $SU(2)$ from each copy of \eref{basicEstring}.  As a result, one obtains
\be \label{Estringtorus}
\begin{tikzpicture}[baseline]
\tikzstyle{every node}=[font=\footnotesize]
\node[draw, circle] (node1) at (-2,-1) {$2$};
\node[draw, circle] (node2) at (2,-1) {$2$};
\node[draw, rectangle] (sqnode) at (0,1) {$8$};
 \draw[transform canvas={yshift=-2.5pt}] (node1) to  node[below] {\blue $D$} node[below, xshift=-0.5cm] {$\red F_D$} node[xshift=-0.5cm] {$ \red \times$} (node2) ;
  \draw[transform canvas={yshift=2.5pt}] (node1) to node[above] {\blue $U$} node[above, xshift=-0.5cm] {$\red F_U$} node[xshift=-0.5cm] {$ \red \times$}  (node2);
\draw[draw=black,solid, -<-=0.5]  (node1) to node[left] {\blue $L$} (sqnode);
\draw[draw=black,solid, -<-=0.5]  (sqnode) to node[right] {\blue $R$} (node2);
\end{tikzpicture}
\ee
with the superpotential
\be
W= L U R +LDR + F_U(UU) + F_D (D D) ~.
\ee
The index of this theory was computed in \cite[(3.3)]{Kim:2017toz}, where it can be written in terms of characters of $E_7 \times U(1)$ representations.   

In fact, a plethora of 4d SCFTs with interesting IR properties, including enhancement of flavour symmetry, can be obtained by compactifying various 6d theories on a torus or a more general Riemann surface, see \eg~ \cite{Bah:2017gph, Razamat:2017wsk, Kim:2018bpg, Kim:2018lfo, Razamat:2018gro, Razamat:2018gbu, Zafrir:2018hkr, Pasquetti:2019hxf, Sela:2019nqa}. 

\subsection*{Modifying the theories}
An interesting question that could be asked is whether it is possible to glue together the basic building block \eref{basic5dwall} in a similar fashion as described above in order to obtain a theory analogous to \eref{Estringtorus}; for example, for $N=2$, we have
\be \label{naivemodify}
\begin{tikzpicture}[baseline]
\tikzstyle{every node}=[font=\footnotesize]
\node[draw, circle] (node1) at (-2,-1) {$2$};
\node[draw, circle] (node2) at (2,-1) {$2$};
\node[draw, rectangle] (sqnode) at (0,1) {$4$};
 \draw[transform canvas={yshift=-2.5pt}] (node1) to  node[below] {\blue $D$} node[below, xshift=-0.5cm] {$\red F_D$} node[xshift=-0.5cm] {$ \red \times$} (node2) ;
  \draw[transform canvas={yshift=2.5pt}] (node1) to node[above] {\blue $U$} node[above, xshift=-0.5cm] {$\red F_U$} node[xshift=-0.5cm] {$ \red \times$}  (node2);
\draw[draw=black,solid, -<-=0.5]  (node1) to node[left] {\blue $L$} (sqnode);
\draw[draw=black,solid, -<-=0.5]  (sqnode) to node[right] {\blue $R$} (node2);
\node[draw=none] at (0,-2) {$W= L U R +LDR + F_U(UU) + F_D (D D)$};
\end{tikzpicture}
\ee
We emphasize that the crucial difference between \eref{naivemodify} and \eref{Estringtorus} is that the 5d gauge theory associated with the former has a UV completion in 5d, whereas that associated with the latter has a UV completion in 6d.  Therefore, \eref{Estringtorus} has a natural interpretation as coming from the compactification of the 6d theory on a torus, which can be obtained by closing the tube, whereas \eref{naivemodify} does not.  In fact, the superpotential and the condition for the non-anomalous $R$-symmetry fixes the $R$-charges of $(U, D, L, R, F_U,F_D)$ to be $(0,0,1,1,2,2)$.  At this stage, we should further introduce the flipping field $F_{UD}$ together with superpotential $F_{UD}(UD)$ that flips the operator $UD$, which falls below the unitarity bound. This leads to the conformal anomalies $(a,c)=\left(\frac{3}{16},\frac{1}{8} \right)$, which implies that the theory flows to the theory of a free vector multiplet.  This implies that such a simple and naive modification of \eref{Estringtorus} to \eref{naivemodify} does not lead to an interesting interacting SCFT.

This, on the other hand, suggests that the superpotential we turned on in \eref{naivemodify} is too restrictive.  We may further modify the theory by dropping the term $LDR$ and the flipping field $F_U$ and consider instead the following theory 
\be
\begin{tikzpicture}[baseline]
\tikzstyle{every node}=[font=\footnotesize]
\node[draw, circle] (node1) at (-2,-1) {$2$};
\node[draw, circle] (node2) at (2,-1) {$2$};
\node[draw, rectangle] (sqnode) at (0,1) {$4$};
 \draw[transform canvas={yshift=-2.5pt}] (node1) to  node[below] {\blue $D$} node[below, xshift=-0.5cm] {$\red F_D$} node[xshift=-0.5cm] {$ \red \times$} (node2) ;
  \draw[transform canvas={yshift=2.5pt}] (node1) to node[above] {\blue $U$}  (node2);
\draw[draw=black,solid, -<-=0.5]  (node1) to node[left] {\blue $L$} (sqnode);
\draw[draw=black,solid, -<-=0.5]  (sqnode) to node[right] {\blue $R$} (node2);
\node[draw=none] at (0,-2) {$W= LUR + F_D(DD) $};
\end{tikzpicture}
\ee
As it will be discussed in section \eref{sec:enhanceSU2w4}, this theory turns out to flow to a decoupled free chiral multiplet, which is identified with the operator $UD$, together with a 4d $\CN=2$ SCFT, described by the 4d $\CN=2$ $SU(2)$ gauge theory with four flavours of fundamental hypermultiplets.  The latter has an $SO(8)$ flavour symmetry. We see that not only the flavour symmetry gets enhanced from $SU(4) \times U(1)$ to $SO(8)$, but supersymmetry also gets enhanced from $\CN=1$ to $\CN=2$.  

This naturally leads to a question whether we can obtain more 4d $\CN=1$ gauge theories with interesting IR properties by modifying the quivers in a similar way as described above. The main objective of this paper is to construct and study a number of such theories.   Our approach is as follows.  We start with 4d $\CN=1$ gauge theories arising from compactification of 6d SCFTs on a torus with fluxes, discussed in \cite{Kim:2017toz, Kim:2018bpg, Kim:2018lfo, Zafrir:2018hkr, Pasquetti:2019hxf}.  The theories are then modified by (1) reducing the number of flavours if this is allowed by gauge anomaly cancellation, (2) dropping some superpotential terms, and (3) adding or dropping flipping fields. As a result, we find several theories that flow to SCFTs with enhanced flavour symmetry, and possibly with enhanced supersymmetry in some cases. Note that as a result of step (1), it is tempting to regard the resulting theory as being obtained by gluing together certain basic building blocks that are associated with duality walls of some 5d gauge theory whose UV completion is in 5d \cite{Gaiotto:2015una}, instead of 6d. However, while these theories are inspired by theories related to 5d domain wall theories, in this paper we do not explicitly study the theories living on the 5d domain walls. The theories studied in this paper were mostly chosen by the existence of interesting IR dynamics, and may or may not have an higher dimensional interpretation. We reserve a more in-depth study of such an interpretation to future work.  


\subsection*{Organization of the paper}
The paper is organized as follows. In section \ref{sec:enhanceSUNp1}, we propose a 4d $\CN=1$ gauge theory that flows to the 4d $\CN=2$ $SU(N+1)$ gauge theory with $2N+2$ flavours of fundamental hypermultiplets and a decoupled free chiral multiplet.  In section \ref{sec:A1D4}, a 4d $\CN=1$ gauge theory that flows to the $(A_1, D_4)$ Argyres--Douglas SCFT is investigated.  This theory turns out to be Seiberg dual to the theory proposed in \cite{Agarwal:2016pjo}.  In section \ref{sec:modD5conf}, we consider modifications of quivers from the minimal $(D_5,D_5)$ conformal matter on a torus with fluxes.  In particular, we discuss a 4d $\CN=1$ gauge theory that flows to the 4d $\CN = 2$ $SO(4)$ gauge theory with $2$ flavours of hypermultiplets in the vector representation.  In section \ref{sec:EUSp2N}, we study a 4d $\CN=1$ quiver gauge theory containing an SCFT known as $E[USp(2N)]$, which was first proposed in \cite{Pasquetti:2019hxf} and is reviewed in appendix \ref{euspapp}, as a component. We discuss the enhancement of the flavour symmetry in the IR.  In section \ref{sec:SU9enhancement}, we study a quiver theory with the $USp(4) \times SU(3)$ gauge group that is a modification of the $(D_5,D_5)$ conformal matter on a torus with fluxes \cite{Kim:2018bpg, Zafrir:2018hkr}.  For the model that we propose, it is found that the flavour symmetry gets enhanced in the IR.  We also discuss a subtlety regarding the accidental symmetry of this model.  We then conclude the paper in section \ref{sec:conclusion}.  The basic notion of the supersymmetric index of $4d$ $\CN=1$ SCFTs is summarized in appendix \ref{app:index}.

\section{Flowing to the 4d $\CN=2$ $SU(N+1)$ with $2N+2$ flavours} \label{sec:enhanceSUNp1}
In this section, we consider a 4d $\CN=1$ quiver gauge theory that flows to the $\CN=2$ $SU(N+1)$ gauge theory with $2N+2$ flavours of fundamental hypermultiplets.  We start by exploring the case of $N=1$ and then move on to the case of general $N$.

\subsection{The case of $N=1$} \label{sec:enhanceSU2w4}
Let us consider the following theory:
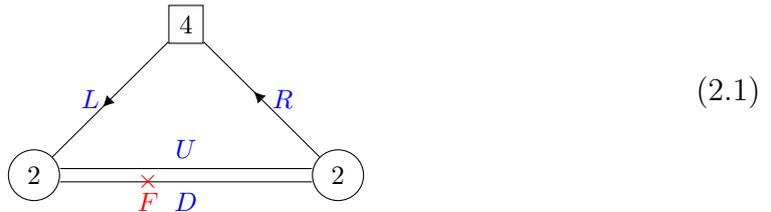
\be \label{model12meq4}
\begin{tikzpicture}[baseline]
\tikzstyle{every node}=[font=\footnotesize]
\node[draw, circle] (node1) at (-2,-1) {$2$};
\node[draw, circle] (node2) at (2,-1) {$2$};
\node[draw, rectangle] (sqnode) at (0,1) {$4$};
 \draw[transform canvas={yshift=-2.5pt}] (node1) to  node[below] {\blue $D$} node[below, xshift=-0.5cm] {$\red F$} node[xshift=-0.5cm] {$ \red \times$} (node2) ;
  \draw[transform canvas={yshift=2.5pt}] (node1) to node[above] {\blue $U$} (node2);
\draw[draw=black,solid, -<-=0.5]  (node1) to node[left] {\blue $L$} (sqnode);
\draw[draw=black,solid, -<-=0.5]  (sqnode) to node[right] {\blue $R$} (node2);
\end{tikzpicture}
\ee
with the superpotential
\be
W= L U R + F (D D)~.
\ee
where $F$ is the flipping field for the gauge invariant quantitiy $D D \equiv \epsilon_{\alpha \beta} \epsilon^{\alpha' \beta'} (D)^\alpha_{\alpha'}  (D)^\beta_{\beta'}$, with $\alpha,\beta=1,2$ the indices for the left gauge group, and $\alpha', \beta'=1,2$ the indices for the right gauge group. This is a modification of the rank-one $E$-string theory on a torus with a flux that breaks $E_8$ to $E_7\times U(1)$ \cite[fig. 3]{Kim:2017toz}.  In comparison with that reference, we lower the number of flavours from $8$ to $4$, drop the flipping field for $UU$, and drop the superpotential term $LDR$.

The superpotential and the condition for the non-anomalous symmetry imply that this theory has one non-anomalous $U(1)$ flavour symmetry, whose fugacity is denoted by $d$.  
The superconformal $R$-charges of the chiral fields can be determined using $a$-maximisation \cite{Intriligator:2003jj}.  We summarize these charges in the following diagram
\be \label{chargemodel1}
\begin{tikzpicture}[baseline]
\tikzstyle{every node}=[font=\footnotesize]
\node[draw, circle] (node1) at (-2,-1) {$2$};
\node[draw, circle] (node2) at (2,-1) {$2$};
\node[draw, rectangle] (sqnode) at (0,1) {$4$};
 \draw[transform canvas={yshift=-2.5pt}] (node1) to  node[below] {\blue $t^{0} d^0$} node[below, xshift=-0.5cm] {} node[xshift=-0.5cm] {$ \red \times$} (node2) ;
  \draw[transform canvas={yshift=2.5pt}] (node1) to node[above] {\blue $t^{\frac{2}{3}} d^2$} (node2);
\draw[draw=black,solid, -<-=0.5]  (node1) to node[left] {\blue $t^{\frac{2}{3}} d^{-1}$} (sqnode);
\draw[draw=black,solid, -<-=0.5]  (sqnode) to node[right] {\blue $t^{\frac{2}{3}} d^{-1}$} (node2);
\end{tikzpicture}
\ee
where the powers of the fugacity $t$ denote the \emph{exact} superconformal $R$-charges.  The conformal anomalies are
\be \label{acmodel1}
(a,c) = \left( \frac{47}{48}, \frac{29}{24} \right)~.
\ee
Observe that the gauge invariant quantity $U D$ has $R$-charge $\frac{2}{3}$ and is therefore a free field, which decouples. Subtracting the conformal anomalies of a free chiral multiplet, $(a,c)_{\text{free chiral}} = (\frac{1}{48}, \frac{1}{24})$, from \eref{acmodel1}, we obtain
\be
(a',c') =  \left( \frac{47}{48} - \frac{1}{48}, \frac{29}{24} - \frac{1}{24}\right)  = \left( \frac{23}{24},\frac{7}{6} \right)~.
\ee
This turns out to be the conformal anomalies of 4d $\CN=2$ $SU(2)$ gauge theory with $4$ flavours.  In particular, this suggests that supersymmetry gets enhanced in the IR.

Let us compute the index of \ref{model12meq4}, whose details are collected in appendix \eref{sec:model12meq4}.  After factoring out the contribution from the free chiral multiplet (which can be achieved, for example, by flipping $UD$) we obtain
\bes{ \label{indexmodel1}
&1+ \left[d^4 + {\blue d^{-2} \left( \chi^{SU(4)}_{[1,0,1]} (\vec u) + 2\chi^{SU(4)}_{[0,1,0]} ( \vec u)+1   \right) } \right] t^{\frac{4}{3}} \\
& \quad - d^2 (y+y^{-1}) t^{\frac{5}{3}} + \left[ - \chi^{SU(4)}_{[1,0,1]} (\vec u) - 2\chi^{SU(4)}_{[0,1,0]} ( \vec u)-1 \right] t^2 +\ldots~.
}
where $\vec u = (u_1,u_2,u_3)$ denotes the $SU(4)$ fugacities corresponding to the square node in quiver \eref{model12meq4}. This can be compared with the index of the $\CN=2$ $SU(2)$ gauge theory with $4$ flavours, whose $SO(8)$ flavour symmetry is decomposed into a subgroup $SU(4) \times U(1)_b$:
\bes{
&1+ \left[d^4 + {\blue d^{-2} \left( \chi^{SU(4)}_{[1,0,1]} (\vec u) + (b^2+b^{-2}) \chi^{SU(4)}_{[0,1,0]} ( \vec u)+1   \right) } \right] t^{\frac{4}{3}} \\
& \quad - d^2 (y+y^{-1}) t^{\frac{5}{3}} + \left[-  \chi^{SU(4)}_{[1,0,1]} (\vec u) -(b^2+b^{-2})\chi^{SU(4)}_{[0,1,0]} ( \vec u)-1 \right] t^2 +\ldots~.
}
The blue terms correspond to the moment map operators transforming under the adjoint representation of $SO(8)$, written in terms of representations of $SU(4) \times U(1)_b$; these operators are mapped to the gauge invariant combinations $LDR$ in \eref{model12meq4}.  The term $d^4 t^{\frac{4}{3}}$ corresponds to the Coulomb branch operator; this is mapped to $U^2$ in \eref{model12meq4}.  Here the $SU(2)\times U(1)$ $R$-symmetry of the $\CN=2$ theory is decomposed into a subgroup $U(1)_R \times U(1)_d$ symmetry, where $U(1)_R$ is the $R$-symmetry of the $\CN=1$ theory and $U(1)_d$ commutes with $U(1)_R$.  The fugacity $b$ corresponds to the baryonic symmetry of the $\CN=2$ theory.  This is not manifest in the description \eref{chargemodel1} of the $\CN=1$ theory but is emergent in the IR.  This is the reason why we cannot refine the index \eref{indexmodel1}, which was computed using \eref{chargemodel1}, with respect to the fugacity $b$.

Finally, we note that it is possible to understand and motivate this result as follows. First, we note from figure \eref{chargemodel1} that the field $D$ has zero charges under all global symmetries and so there is no impediment to it acquiring a vev. Therefore, under the usual way of thought in quantum field theory, we expect this field to acquire a vev dynamically during the RG flow. The effect of this vev should be to identify the two $SU(2)$ gauge groups, leading to only a single $SU(2)$ gauge group, the diagonal one. The additional vector multiplets are Higgsed together with most of the components of the bifundamental $D$. The bifundamental $U$, becomes a field in the adjoint representation of the remaining $SU(2)$ and a singlet chiral field. The superpotential $LUR$ then couples the adjoint field with the fields $L$ and $R$. Overall, we end up precisely with the $\CN=2$ $SU(2)$ gauge theory with $4$ flavours, plus a single free chiral field that can be identified with the gauge invariant given by $U^2$.

\subsection{General $N$}
An interesting generalization of \eref{model12meq4} is to consider the following model:
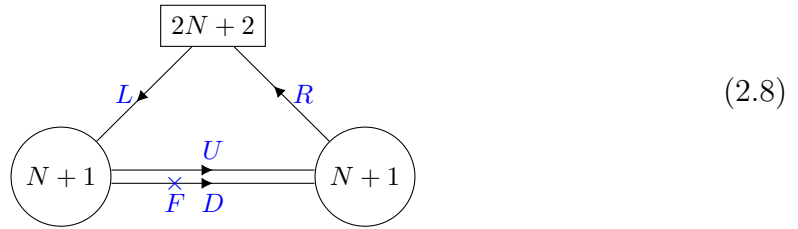
\be \label{model1genN}
\begin{tikzpicture}[baseline]
\tikzstyle{every node}=[font=\footnotesize]
\node[draw, circle] (node1) at (-2,-1) {$N+1$};
\node[draw, circle] (node2) at (2,-1) {$N+1$};
\node[draw, rectangle] (sqnode) at (0,1) {$2N+2$};
 \draw[transform canvas={yshift=-2.5pt},->-=0.5] (node1) to  node[below] {\blue $D$} node[below, xshift=-0.5cm] {$\blue F$} node[xshift=-0.5cm] {$ \blue \times$} (node2) ;
  \draw[transform canvas={yshift=2.5pt},->-=0.5] (node1) to node[above] {\blue $U$} (node2);
\draw[draw=black,solid, -<-=0.5]  (node1) to node[left] {\blue $L$} (sqnode);
\draw[draw=black,solid, -<-=0.5]  (sqnode) to node[right] {\blue $R$} (node2);
\end{tikzpicture}
\ee
with superpotential
\be
W= LDR+ F D^{N+1} ~.
\ee

This model can also be thought of as a modification of a 4d theory descending from the compactification of a 6d $(1,0)$ SCFT, similarly to the previous model. Here the 4d theory in question is the one in \cite[fig. 7]{Kim:2018bpg}, which comes from a compactification of the 6d $(1,0)$ SCFT known as the $(D_{N+3},D_{N+3})$ conformal matter \cite{DelZotto:2014hpa}. Like in the previous case, the 4d theory in \cite{Kim:2018bpg} is based on 5d domain walls between different 5d gauge theory descriptions of the 6d SCFT on the circle. In line with our general approach here, the modification in \eqref{model1genN} then corresponds to changing the 5d matter content by the removal of fundamental fields such that the 5d gauge theory now has a 5d SCFT as its UV completion\footnote{The $(D_{N+3},D_{N+3})$ conformal matter on the circle for $N>1$ has several different 5d gauge theory description, leading to multiple interesting domain wall theories. However, not all cases support a generalization to a smaller number of flavors, while others are more intricate making the calculation we wish to perform involved for generic $N$. We shall return to consider cases based on other domain walls between 5d gauge theory descriptions for the $(D_{5},D_{5})$ conformal matter theory in section \ref{sec:modD5conf}.}. Nevertheless, this does not guarantee that the theory in figure \eqref{model1genN} has an interesting higher dimensional origin as it may not be a domain wall theory associated with the modified 5d gauge theory and its associated 5d SCFT.

In the same way as \eref{chargemodel1}, this theory has one non-anomalous $U(1)$ flavour symmetry, whose fugacity is denoted by $d$.  The $U(1)_d$ charges and superconformal $R$-charges of each chiral field are depicted in the following diagram:
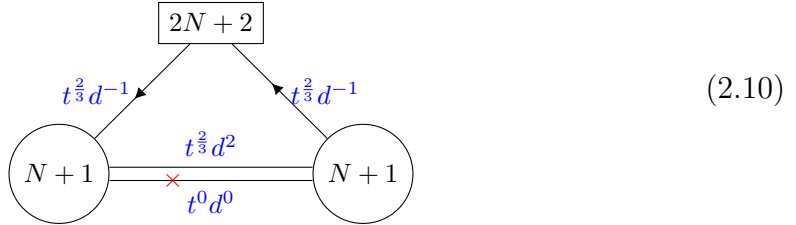
\be \label{chargemodel1genN}
\begin{tikzpicture}[baseline]
\tikzstyle{every node}=[font=\footnotesize]
\node[draw, circle] (node1) at (-2,-1) {$N+1$};
\node[draw, circle] (node2) at (2,-1) {$N+1$};
\node[draw, rectangle] (sqnode) at (0,1) {$2N+2$};
 \draw[transform canvas={yshift=-2.5pt}] (node1) to  node[below] {\blue $t^{0} d^0$} node[below, xshift=-0.5cm] {} node[xshift=-0.5cm] {$ \red \times$} (node2) ;
  \draw[transform canvas={yshift=2.5pt}] (node1) to node[above] {\blue $t^{\frac{2}{3}} d^2$} (node2);
\draw[draw=black,solid, -<-=0.5]  (node1) to node[left] {\blue $t^{\frac{2}{3}} d^{-1}$} (sqnode);
\draw[draw=black,solid, -<-=0.5]  (sqnode) to node[right] {\blue $t^{\frac{2}{3}} d^{-1}$} (node2);
\end{tikzpicture}
\ee
The conformal anomalies are
\be
(a,c) = \left( \frac{1}{48} \left(14 N^2+28 N+5\right),\frac{1}{24} \left(8 N^2+16 N+5\right)\right)~.
\ee
Similarly to \eref{model12meq4}, we see that the gauge invariant combination $UD^{N}$ has $R$-charge $2/3$ and is therefore free and decouples.  Upon subtracting $(a,c)_{\text{free chiral}}=  \left(\frac{1}{48}, \frac{1}{24} \right)$, we obtain
\bes{
(a',c') &=  \left( \frac{1}{48} \left(14 N^2+28 N+5\right),\frac{1}{24} \left(8 N^2+16 N+5\right)\right) - \left(\frac{1}{48}, \frac{1}{24} \right) \\
&=\left( \frac{1}{24} \left(7 N^2+14 N+2\right),\frac{1}{6} \left(2 N^2+4 N+1\right) \right)~.
}
This turns out to be precisely the conformal anomalies for 4d $\CN=2$ $SU(N+1)$ gauge theory with $2N+2$ flavours.

We compute the index of \eref{model1genN} for $N=2$ and obtain
\bes{
&1+ \left[d^4 + {\blue d^{-2} \left( \chi^{SU(6)}_{[1,0,0,0,1]} (\vec u)+1   \right) } \right] t^{\frac{4}{3}} - d^2 (y+y^{-1}) t^{\frac{5}{3}}  \\
& \quad + \left[ - \chi^{SU(6)}_{[1,0,0,0,1]} (\vec u) -1 +{\purple 2d^{-3}\chi^{SU(6)}_{[0,0,1,0,0]} ( \vec u)}+ d^6 \right] t^2 +\ldots~.
}
This can be compared with the index for the $\CN=2$ $SU(3)$ gauge theory with $6$ flavours:
\bes{
&1+ \left[d^4 + {\blue d^{-2} \left( \chi^{SU(6)}_{[1,0,0,0,1]} (\vec u)+1   \right) } \right] t^{\frac{4}{3}} - d^2 (y+y^{-1}) t^{\frac{5}{3}}  \\
& \quad + \left[ - \chi^{SU(6)}_{[1,0,0,0,1]} (\vec u) -1 + {\purple d^{-3}(b^3+b^{-3})\chi^{SU(6)}_{[0,0,1,0,0]} ( \vec u)}+ d^6 \right] t^2 +\ldots~.
}
where $b$ is the fugacity for the baryonic symmetry $U(1)_b$  of the $\CN=2$ theory.  This symmetry is not manifest in the description \eref{chargemodel1genN} of the $\CN=1$ theory, but is emergent in the IR.
Similarly to the $N=1$ case, the $U(1)_d$ symmetry is the commutant of the $\CN=1$ $R$-symmetry in the $\CN=2$ $SU(2) \times U(1)$ $R$-symmetry.  The blue terms correspond to the moment map operators in the adjoint representation of $SU(6) \times U(1)_b$; these are mapped to the gauge invariant combinations $LDR$ in \eref{model1genN}.   The term $d^4 t^{\frac{4}{3}}$ denotes the Coulomb branch operator $\tr(\phi^2)$, where $\phi$ is the complex scalar in the $\CN=2$ vector multiplet; this operator is mapped to $U^2D$ in \eref{model1genN}.   The marginal operators are represented by the positive terms at order $t^2$, and they are as follows.  The brown terms correspond to the baryons and antibaryons in the $\CN=2$ theory; they are mapped to $L^3$ and $R^3$ in \eref{model1genN}.  The term $d^6 t^2$ corresponds to the Coulomb branch operator $\tr (\phi^3)$ of the $\CN=2$ theory; it is mapped to the operator $U^3$ in \eref{model1genN}.  The negative terms at order $t^2$ confirm that the non-$R$ global symmetry of the theory is indeed $SU(6) \times U(1)_b$ \footnote{The contribution of the conserved current for $U(1)_d$ is canceled against the contribution of the $\CN=2$ preserving marginal operator that is associated with the gauge coupling. Hence, both are absent in the index.}.

Like in the $N=1$ case, we can understand and motivate this result as the field $D$ has zero charges under all global symmetries and so there is no impediment to it acquiring a vev. Therefore, we again expect such a vev to be dynamically generated, leading to the identification of the two $SU(N+1)$ groups and the collapse of the quiver to a single $SU(N+1)$ gauge theory. Following what happens to the matter content, we again see that we just get the $\CN=2$ $SU(N+1)$ gauge theory with $2N+2$ fundamental flavours, plus a single free chiral field.

\section{Flowing to the $(A_1, D_4)$ Argyres-Douglas theory} \label{sec:A1D4}
Let us now consider the following theory:
\be \label{model2}
\begin{tikzpicture}[baseline]
\tikzstyle{every node}=[font=\footnotesize]
\node[draw, circle] (node1) at (-2,-1) {$2$};
\node[draw, circle] (node2) at (2,-1) {$2$};
\node[draw, rectangle] (sqnode) at (0,1) {$2$};
\node[draw, rectangle] (sqnode2) at (4,-1) {$2$};
 \draw[transform canvas={yshift=-2.8pt}] (node1) to  node[below] {\blue $D$} node[below, xshift=-0.6cm] {$\red F_D$} node[xshift=-0.6cm] {$ \red \times$} (node2) ;
  \draw[transform canvas={yshift=2.8pt}] (node1) to node[above] {\blue $U$} node[above, xshift=-0.6cm] {$\red F_U$} node[xshift=-0.6cm] {$ \red \times$} (node2);
\draw[draw=black,solid]  (node1) to node[left] {\blue $L$} node[near start,left] {$\red F_L$} node[near start] {$ \red \times$} (sqnode);
\draw[draw=black,solid]  (sqnode) to node[right] {\blue $R$} (node2);
\draw[draw=black,solid]  (node2) to node[above] {\blue $Q$} node[above, xshift=-0.4cm] {$\red F_Q$} node[xshift=-0.4cm] {$ \red \times$} (sqnode2);
\node[draw=none]  (cross1) at (0.7,-1) {\Large \red $\times$}node at (0.7,-1.4) {\red $F_1$};
\node[draw=none]  (cross2) at (1.2,-1) {\huge \red $\times$}node at (1.2,-1.4) {\red $F_2$};
\end{tikzpicture}
\ee
and turn on the superpotential:
\bes{ \label{supmodel2}
W&= L U R +(U D)  (L D )^2+ F_U( U U)  \\
& \quad + F_D( D D) + F_L (L L) +F_Q( Q Q)+ F_1 \tr (U D) + F_2 \tr((U D)^2)~.
}
This is again the modification of  the rank-one $E$-string theory on a torus with a flux that breaks $E_8$ to $SO(14)\times U(1)$ \cite[figure 12]{Kim:2017toz}.

The superpotential and the condition for non-anomalous $R$-symmetry imply that there is one non-anomalous $U(1)$ flavour symmetry, whose fugacity is denoted by $d$.  The $U(1)_d$ charges and superconformal $R$-charges of each chiral field are depicted in the following diagram:
\be \label{chargemodel2}
\begin{tikzpicture}[baseline]
\tikzstyle{every node}=[font=\footnotesize]
\node[draw, circle] (node1) at (-2,-1) {$2$};
\node[draw, circle] (node2) at (2,-1) {$2$};
\node[draw, rectangle] (sqnode) at (0,1) {$2$};
\node[draw, rectangle] (sqnode2) at (4,-1) {$2$};
 \draw[transform canvas={yshift=-2.8pt}] (node1) to  node[below] {\blue $t^{\frac{1}{6}} d$} node[below, xshift=-0.6cm] {$$} node[xshift=-0.6cm] {$ \red \times$} (node2) ;
  \draw[transform canvas={yshift=2.8pt}] (node1) to node[above] {\blue $t^{\frac{1}{6}} d$} node[above, xshift=-0.6cm] {$$} node[xshift=-0.6cm] {$ \red \times$} (node2);
\draw[draw=black,solid]  (node1) to node[left, near end] {\blue $t^{\frac{2}{3}} d^{-2}$} node[near start,left] {$$} node[near start, rotate=30] {$ \red \times$} (sqnode);
\draw[draw=black,solid]  (sqnode) to node[right, near start] {\blue $t^{\frac{7}{6}} d$} (node2);
\draw[draw=black,solid]  (node2) to node[above, near end] {\blue $t^{\frac{1}{2}} d^{-3}$} node[above, xshift=-0.4cm] {$$} node[xshift=-0.4cm] {$ \red \times$} (sqnode2);
\node[draw=none]  (cross1) at (0.7,-1) {\Large \red $\times$}node at (0.7,-1.4) {};
\node[draw=none]  (cross2) at (1.2,-1) {\huge \red $\times$}node at (1.2,-1.4) {};
\end{tikzpicture}
\ee
where the powers of the fugacity $t$ denote the \emph{exact} superconformal $R$-charges.
The conformal anomalies are
\be
(a,c) = \left( \frac{7}{12}, \frac{2}{3} \right)~.
\ee
This turns out to be those of the $(A_1, D_4)$ or $H_2$ Argyres--Douglas theory.  In order to see the relation between \eref{model2} and the $(A_1,D_4)$ theory, it is more convenient to apply Seiberg duality \cite{Seiberg:1994pq} to the lower left $SU(2)$ gauge node.


\subsection{Seiberg dual of theory \eref{model2}} \label{sec:seibergmodel2}
Let us apply the Seiberg duality \cite{Seiberg:1994pq} (see also the Intriligator-Pouliot duality \cite{Intriligator:1995ne}) to the lower left $SU(2)$ gauge node in \eref{model2}, which has six fundamental chiral fields (3 flavours) transforming under it.  As a result, we obtain a Wess--Zumino model with $15$ singlets transforming under the rank two antisymmetric representation of the $SU(6)$ acting on the six fundamental chirals. In the quiver theory we do not have the $SU(6)$ as part of it is gauged by the right $SU(2)$ gauge group, and so we should split these $15$ singlets into representations of the $SU(2)$ gauge group and its commutant. Specifically, this gives $4$ mesons $M_{U} = L U$ and $4$ mesons $M_{D} = L D$, both transforming in the bifundamental of $SU(2)$ gauge and the upper global $SU(2)$, $1$ baryon $L^2$, $1$ baryon $U^2$, and 1 baryon $D^2$, which are singlets, and the $4$ fields $U D$.  The latter can be split into the trace part $\tr(U D)$ and the traceless part $X$; in other words, $\tr(X) =0$ and
\be
U D = X + \frac{1}{2} \tr(U D) \mathbf{1}_{2\times 2}~.
\ee
The field $X$ then is a chiral field in the adjoint of the gauge $SU(2)$, while $\tr(U D)$ becomes a singlet chiral field. From the superpotential \eref{supmodel2}, all of the baryons and the trace $\tr(U D)$ are flipped, so they are set to zero in the chiral ring.
We then obtain the following dual theory
\be
\begin{tikzpicture}[baseline]
\tikzstyle{every node}=[font=\footnotesize]
\node[draw, rectangle] (sqnodeL) at (-4,0) {$2$};
\node[draw, circle] (node) at (0,0) {$2$};
\node[draw, rectangle] (sqnodeR) at (4,0) {$2$};
\draw[transform canvas={yshift=0pt}] (sqnodeL) to [bend left=40] node[above]{\blue $M_{U}$} (node);
\draw[transform canvas={yshift=0pt}] (sqnodeL) to node[above] {\blue $R$}(node);
\draw[transform canvas={yshift=0pt}]  (sqnodeL) to [bend right=40] node[below]{\blue $M_{D}$} (node);
\draw[transform canvas={yshift=0pt}]  (node) to node[above]{\blue $Q$} node[above, xshift=-0.6cm] {$\red F_Q$} node[xshift=-0.6cm] {$ \red \times$} (sqnodeR);
\draw[black,solid] (node) edge [out=45,in=135,loop,looseness=8] node[above]{\blue $X$} node[below, yshift=0cm] {$\red F_X$} node {$ \red \times$}  (node);
\end{tikzpicture}
\ee
The superpotential \eref{supmodel2} of the original theory contains the term $U L R \rightarrow M_{U} R$.  This implies that the fields $R$ and $M_{U}$ acquire a mass and can be integrated out. 
We are thus left with the following theory
\be \label{model2dualB}
\begin{tikzpicture}[baseline]
\tikzstyle{every node}=[font=\footnotesize]
\node[draw, rectangle] (sqnodeL) at (-4,0) {$2$};
\node[draw, circle] (node) at (0,0) {$2$};
\node[draw, rectangle] (sqnodeR) at (4,0) {$2$};
\draw[transform canvas={yshift=0pt}]  (sqnodeL) to node[above]{\blue $M_{D}$} (node);
\draw[transform canvas={yshift=0pt}]  (node) to node[above]{\blue $Q$} node[above, xshift=-0.6cm] {$\red F_Q$} node[xshift=-0.6cm] {$ \red \times$} (sqnodeR);
\draw[black,solid] (node) edge [out=45,in=135,loop,looseness=8] node[above]{\blue $X$} node[below, yshift=0cm] {$\red F_X$} node {$ \red \times$}  (node);
\end{tikzpicture}
\ee
The superpotential of this theory can be determined by putting all of the possible gauge and flavour invariants that map to the combinations of the fields in \eref{chargemodel2} with $R$-charge $2$ and $U(1)_d$ charge $0$:
\be  \label{supmodel2dualB}
W = F_X \tr(X^2)+ F_Q (QQ) + X M_D M_D~.
\ee
This theory was in fact studied in section 3.2 of \cite{Agarwal:2016pjo} and section 2.1 of \cite{Benvenuti:2017bpg}.  The last term in the superpotential breaks the $SU(2)$ flavour symmetry corresponding to the left square node to $SO(2) \cong U(1)$.  This, together with the $SU(2)$ corresponding to the right square node, gets enhanced to $SU(3)$ in the IR.  There is also a non-anomalous $U(1)$ symmetry, which can be identified with $U(1)_d$ of the original theory.  The superconformal $R$-charges and $U(1)_d$ charges of the chiral fields are summarised as follows:
\be \label{chargedualmodel2}
\begin{tikzpicture}[baseline]
\tikzstyle{every node}=[font=\footnotesize]
\node[draw, rectangle] (sqnodeL) at (-4,0) {$2$};
\node[draw, circle] (node) at (0,0) {$2$};
\node[draw, rectangle] (sqnodeR) at (4,0) {$2$};
\draw[transform canvas={yshift=0pt}]  (sqnodeL) to node[above]{\blue $t^{\frac{5}{6}} d^{-1}$} (node);
\draw[transform canvas={yshift=0pt}]  (node) to node[above, near end]{\blue $t^{\frac{1}{2}} d^{-3}$} node[above, xshift=-0.6cm] {} node[xshift=-0.6cm] {$ \red \times$} (sqnodeR);
\draw[black,solid] (node) edge [out=45,in=135,loop,looseness=8] node[above]{\blue $t^{\frac{1}{3}} d^2$} node[below, yshift=0cm] {} node {$ \red \times$}  (node);
\end{tikzpicture}
\ee
The conformal anomalies are
\be
(a,c) = \left( \frac{7}{12}, \frac{2}{3} \right)~,
\ee
which are equal to those of the $(A_1, D_4)$ Argyres--Douglas theory, whose index was computed in (5.12) in \cite{Agarwal:2016pjo}.  Using the notation of \eref{chargedualmodel2}, this can be written as\footnote{The notation in  (5.12) in \cite{Agarwal:2016pjo} can be translated to our notation as follows: $\mathfrak{t} = t^{\frac{1}{3}}$ and $v= d^4$.}
\bes{
& 1+d^6 t + \left[d^{-4} {\purple \left(\chi^{SU(2)}_{[2]}(x) + (u+ u^{-1}) \chi^{SU(2)}_{[1]}(x)+1  \right)}  -d^2 (y+y^{-1})\right] t^{\frac{4}{3}} \\
& + d^{-2} t^{\frac{5}{3}}  + \left[ -\left(\chi^{SU(2)}_{[2]}(x) + (u+ u^{-1}) \chi^{SU(2)}_{[1]}(x)+2  \right)  +d^{12}+d^6 (y+y^{-1}) \right] t^2+\ldots~,
}
where $u$ is the $SO(2) \cong U(1)$ fugacity corresponding to the left square node in \eref{chargedualmodel2}, and $x$ is the $SU(2)$ fugacity corresponding to the right square node in \eref{chargedualmodel2}.  The brown terms correspond to the decomposition of the adjoint representation of $SU(3)$ to those of $SU(2)_x \times U(1)_u$.  These are indeed the contributions of the gauge invariant combinations $QXQ$, $M_D Q$ and $F_X$ in \eref{model2dualB}, which are mapped to the moment map operators of the $(A_1,D_4)$ theory, possessing an $SU(3)$ flavour symmetry.  The term $d^{12} t^2$ corresponds to the marginal operator $F_Q^2$.

\section{Modifications of quivers from the minimal $(D_5,D_5)$ conformal matter on a torus with fluxes} \label{sec:modD5conf}
The quivers for 4d theories arising from the compactification of the 6d minimal $(D_5,D_5)$ conformal matter on a torus with various fluxes were presented in figures 29, 30 and 31 of \cite{Kim:2018lfo}.  The idea of constructing such theories was to start from a suitable building block theory corresponding to a sphere with two punctures (\ie~ a cylinder) associated with appropriate 6d flux. Such a flux can be viewed as introducing domain walls in certain 5d gauge theories, whose UV completion is the 6d conformal matter.  Every building block contains an $SU(4) \times SU(4)$ flavour symmetry, which are subgroups of the 6d $SO(20)$ global  symmetry group that were preserved by the fluxes. To form a torus with a given flux, the two punctures of an appropriate cylinder are then glued together.

In this section, we consider a variation of the above 4d theories.  Similarly to the preceding sections, we modify the building block such that the flavour symmetry is $SU(2) \times SU(2)$, instead of $SU(4) \times SU(4)$ as mentioned above. We then glue such building blocks together.  The resulting theories have the same structure as those in figures 29, 30 and 31 of \cite{Kim:2018lfo} but with $SU(4)$ flavour symmetry nodes replaced by $SU(2)$.  The flipping fields and superpotential are then introduced such that the gauge theory has interesting IR properties.

\subsection{A model with an $SU(2)^3 \times U(1)$ flavour symmetry}
We consider the following modification of figure 29 of \cite{Kim:2018lfo}:
\be 
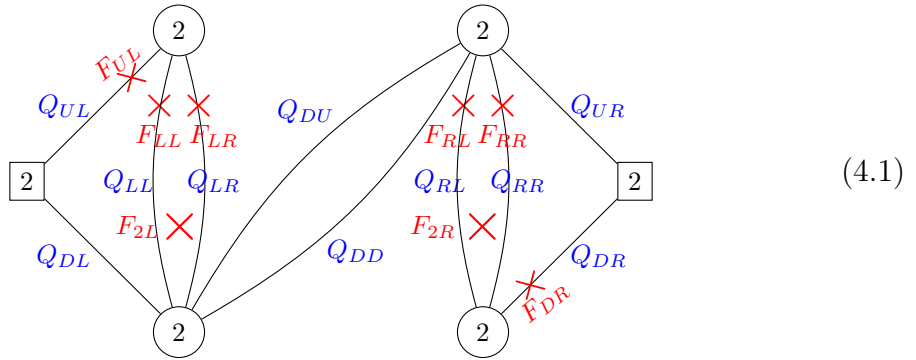
\label{modification29}
\begin{tikzpicture}[baseline]
\tikzstyle{every node}=[font=\footnotesize]
\node[draw, rectangle] (sqL) at (-4,0) {$2$};
\node[draw, circle] (cTL) at (-2,2) {$2$};
\node[draw, circle] (cTR) at (2,2) {$2$};
\node[draw, circle] (cBL) at (-2,-2) {$2$};
\node[draw, circle] (cBR) at (2,-2) {$2$};
\node[draw, rectangle] (sqR) at (4,0) {$2$};
\draw[-] (sqL) to node[left] {\blue $Q_{UL}$}    node[near end, rotate=30]  {\Large \red $\times$} node[near end, rotate=40,above] {\red $F_{UL}$} (cTL);
\draw[-] (sqL) to node[left] {\blue $Q_{DL}$}(cBL);
\draw[-] (sqR) to node[right] {\blue $Q_{UR}$}(cTR);
\draw[-] (sqR) to node[right] {\blue $Q_{DR}$}  node[near end, rotate=30]  {\Large \red $\times$}  node[near end, rotate=40,below] {\red $F_{DR}$} (cBR);
\draw[-] (cTL) to [bend left=15]  node[xshift=-1cm] {\blue $Q_{LL}$} node[xshift=-0.6cm, yshift=1cm]  {\Large \red $\times$} node[xshift=-0.6cm, yshift=0.6cm]  {\red $F_{LL}$} (cBL) ;
\draw[ -] (cTL) to [bend right=15] node[xshift=0.8cm] {\blue $Q_{LR}$} node[xshift=0.6cm, yshift=1cm]  {\Large \red $\times$} node[xshift=0.8cm, yshift=0.6cm]  {\red $F_{LR}$}  node[xshift=0.35cm, yshift=-0.6cm]  {\fontsize{20}{30}\selectfont \red  $\times$} node[xshift=-0.2cm, yshift=-0.6cm]  {\red $F_{2L}$} (cBL); 
\draw[-] (cBL) to  [bend left= 15] node[yshift=0.6cm] {\blue $Q_{DU}$}  (cTR) ;
\draw[-] (cBL) to [bend right=15] node[yshift=-0.6cm]  {\blue $Q_{DD}$}  (cTR);
\draw[-] (cTR) to [bend left=15] node[xshift=-0.9cm] {\blue $Q_{RL}$}  node[xshift=-0.6cm, yshift=1cm]  {\Large \red $\times$}  node[xshift=-0.8cm, yshift=0.6cm]  {\red $F_{RL}$} node[xshift=-0.35cm, yshift=-0.6cm]  {\fontsize{20}{30}\selectfont \red  $\times$} node[xshift=-1cm, yshift=-0.6cm]  {\red $F_{2R}$}   (cBR); 
\draw[-] (cTR) to [bend right=15]  node[xshift=0.8cm] {\blue $Q_{RR}$} node[xshift=0.6cm, yshift=1cm]  {\Large \red $\times$} node[xshift=0.6cm, yshift=0.6cm]  {\red $F_{RR}$}  (cBR) ;
\end{tikzpicture}
\ee
with superpotential
\bes{ \label{supfig29original}
W&= Q_{UL} Q_{LL} Q_{DL} + Q_{UR} Q_{RR} Q_{DR} +Q_{LL}Q_{DU}Q_{DD}Q_{LR}+Q_{RL}Q_{DU}Q_{DD}Q_{RR}\\
& +F_{UL} Q_{UL}^2+ F_{LL} Q_{LL}^2 + F_{LR} Q_{LR}^2 + F_{2L} Q_{LL} Q_{LR} \\  
& +F_{DR} Q_{DR}^2 + F_{RL} Q_{RL}^2  + F_{RR} Q_{RR}^2 + F_{2R} Q_{RL} Q_{RR} ~. 
}
There are two non-anomalous $U(1)$ symmetries whose fugacities are denoted by $d_1$ and $d_2$. Each chiral field in the quiver carries the global charges as indicated in the diagram below:

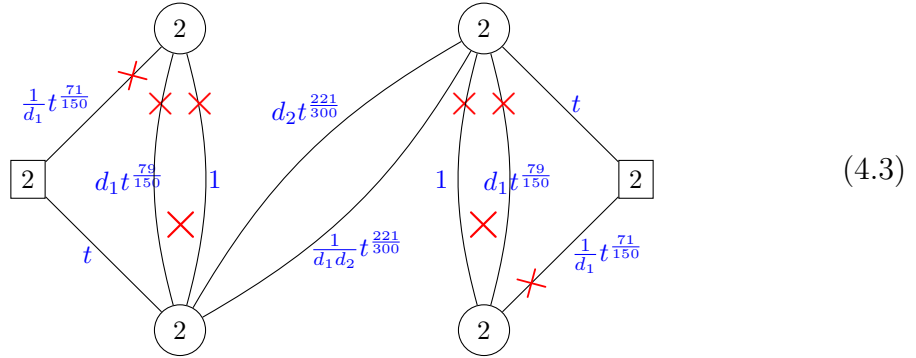
\be
\begin{tikzpicture}[baseline]
\tikzstyle{every node}=[font=\footnotesize]
\node[draw, rectangle] (sqL) at (-4,0) {$2$};
\node[draw, circle] (cTL) at (-2,2) {$2$};
\node[draw, circle] (cTR) at (2,2) {$2$};
\node[draw, circle] (cBL) at (-2,-2) {$2$};
\node[draw, circle] (cBR) at (2,-2) {$2$};
\node[draw, rectangle] (sqR) at (4,0) {$2$};
\draw[-] (sqL) to node[left] {\blue $\frac{1}{d_1} t^{\frac{71}{150}} $}   node[near end, rotate=30]  {\Large \red $\times$}  (cTL);
\draw[-] (sqL) to node[left] {\blue $t$}(cBL);
\draw[-] (sqR) to node[right] {\blue $t$}(cTR);
\draw[-] (sqR) to node[right] {\blue $\frac{1}{d_1} t^{\frac{71}{150}} $} node[near end, rotate=30]  {\Large \red $\times$}  (cBR);
\draw[-] (cTL) to [bend left=15]  node[xshift=-1cm] {\blue $d_1 t^{\frac{79}{150}}$} node[xshift=-0.6cm, yshift=1cm]  {\Large \red $\times$} (cBL) ;
\draw[ -] (cTL) to [bend right=15] node[xshift=0.8cm] {\blue $1$}  node[xshift=0.6cm, yshift=1cm]  {\Large \red $\times$}   node[xshift=0.35cm, yshift=-0.6cm]  {\fontsize{20}{30}\selectfont \red  $\times$}   (cBL); 
\draw[-] (cBL) to  [bend left= 15] node[yshift=0.6cm] {\blue $ d_2 t^{\frac{221}{300}}$}  (cTR) ;
\draw[-] (cBL) to [bend right=15] node[yshift=-0.6cm]  {\blue $ \frac{1}{d_1d_2} t^{\frac{221}{300}}$}  (cTR);
\draw[-] (cTR) to [bend left=15] node[xshift=-0.9cm] {\blue $1$}  node[xshift=-0.6cm, yshift=1cm]  {\Large \red $\times$}  node[xshift=-0.35cm, yshift=-0.6cm]   {\fontsize{20}{30}\selectfont \red  $\times$}  (cBR); 
\draw[-] (cTR) to [bend right=15]  node[xshift=0.8cm] {\blue $d_1 t^{\frac{79}{150}}$} node[xshift=0.6cm, yshift=1cm]  {\Large \red $\times$} (cBR) ;
\end{tikzpicture}
\ee
where the powers of the fugacity $t$ denote the \emph{approximate} superconformal $R$-charges\footnote{The exact mixing coefficients $\alpha$ such that $U(1)_R = \frac{2}{3} + \alpha U(1)_c$ for $c={d_1}, {d_2}$ are $-\left( \frac{7}{3}-\frac{2 \sqrt{14}}{3} \right) \approx \frac{29}{150}$ and $\frac{4}{3}-\frac{\sqrt{14}}{3} \approx \frac{7}{100}$, respectively.}.  The conformal anomalies of this theory are
\be \label{ac29}
(a,c) =\left(7 \sqrt{14}-\frac{99}{4},\hspace{0.2cm} \frac{29}{2} \sqrt{\frac{7}{2}}-\frac{51}{2}\right)~.
\ee

\subsubsection{Seiberg dual of theory \eref{modification29}}
We can Seiberg dualise the top left and bottom right nodes of \eref{modification29}, in a similar way to that described in section \ref{sec:seibergmodel2}.  As a result, we obtain the following quiver
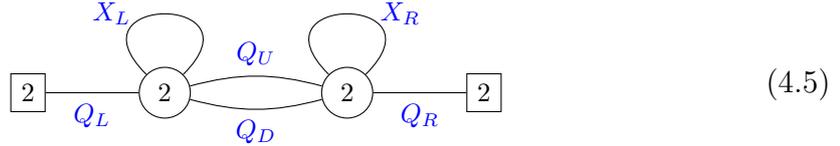
\be \label{dualmodification29}
\begin{tikzpicture}[baseline]
\tikzstyle{every node}=[font=\footnotesize]
\node[draw, rectangle] (sqL) at (-3,0) {$2$};
\node[draw, circle] (cL) at (-1.2,0) {$2$};
\node[draw, circle] (cR) at (1.2,0) {$2$};
\node[draw, rectangle] (sqR) at (3,0) {$2$};
\draw[transform canvas={yshift=0pt}]  (sqL) to node[below]{\blue $Q_L$}  (cL);
\draw[transform canvas={yshift=0pt}]  (cL) to [bend left=15] node[above]{\blue $Q_{U}$}  (cR);
\draw[transform canvas={yshift=0pt}]  (cL) to [bend right=15] node[below]{\blue $Q_{D}$}  (cR);
\draw[black,solid] (cL) edge [out=45,in=135,loop,looseness=8] node[xshift=-0.7cm]{\blue $X_L$}  (cL); 
\draw[black,solid] (cR) edge [out=45,in=135,loop,looseness=8] node[xshift=0.7cm]{\blue $X_R$} (cR); 
\draw[transform canvas={yshift=0pt}]  (cR) to node[below]{\blue $Q_R$}  (sqR);
\end{tikzpicture}
\ee
where the chiral fields of this theory are mapped to the combinations in \eref{modification29} as follows:
\be
\begin{array}{ccc}
 \eref{dualmodification29} && \eref{modification29} \\
Q_L & \quad \longleftrightarrow &\quad  Q_{UL} Q_{LR}  \\
Q_R  & \quad \longleftrightarrow &\quad Q_{DR} Q_{RL}  \\
Q_U & \quad \longleftrightarrow & \quad Q_{DU}\\
Q_D & \quad \longleftrightarrow & \quad Q_{DD}\\
X_L &  \quad \longleftrightarrow & \quad \text{traceless part of  $Q_{LL} Q_{LR}$} \\
X_R &  \quad \longleftrightarrow & \quad \text{traceless part of $Q_{RL} Q_{RR}$} \\
\end{array}
\ee
where we remark that the traces of $Q_{LL} Q_{LR}$ and $Q_{RL} Q_{RR}$ are flipped by $F_{2L}$ and $F_{2R}$ according to \eref{supfig29original}, and so $X_L$ and $X_R$ transform under the adjoint representation of each $SU(2)$ gauge group.

Each chiral field in the dual theory carries the global charges as indicated in the diagram below:
\be
\begin{tikzpicture}[baseline]
\tikzstyle{every node}=[font=\footnotesize]
\node[draw, rectangle] (sqL) at (-3,0) {$2$};
\node[draw, circle] (cL) at (-1.2,0) {$2$};
\node[draw, circle] (cR) at (1.2,0) {$2$};
\node[draw, rectangle] (sqR) at (3,0) {$2$};
\draw[transform canvas={yshift=0pt}]  (sqL) to node[below]{\blue $\frac{1}{d_1}  t^{\frac{71}{150}}$}  (cL);
\draw[transform canvas={yshift=0pt}]  (cL) to [bend left=15] node[above]{\blue $d_2 t^{\frac{221}{300}} $}  (cR);
\draw[transform canvas={yshift=0pt}]  (cL) to [bend right=15] node[below]{\blue $\frac{1}{d_1d_2} t^{\frac{221}{300}} $}  (cR);
\draw[black,solid] (cL) edge [out=45,in=135,loop,looseness=8] node[xshift=-1cm]{\blue ${d_1}  t^{\frac{79}{150}}$}  (cL); 
\draw[black,solid] (cR) edge [out=45,in=135,loop,looseness=8] node[xshift=1cm]{\blue ${d_1}  t^{\frac{79}{150}}$} (cR); 
\draw[transform canvas={yshift=0pt}]  (cR) to node[below]{\blue $\frac{1}{d_1}  t^{\frac{71}{150}}$}  (sqR);
\end{tikzpicture}
\ee
The superpotential of the dual theory can be determined by gauge and flavour invariant combinations in the above quiver that have $R$-charge $2$:
\bes{
W &=  Q_U X_L Q_D  + Q_U X_R Q_D + (Q_L X_L)^2 + (Q_R X_R)^2~.
}
The conformal anomalies of \eref{dualmodification29} are indeed equal to \eref{ac29}, as it should be.

In fact, the $SU(2)$ global symmetry in \eref{modification29} and \eref{dualmodification29} can be made manifest by setting
\be 
d_2 = w d_1^{-\frac{1}{2}}~,
\ee
where $w$ is the $SU(2)$ fugacity. This $SU(2)$ is just the one rotating the two $SU(2)\times SU(2)$ bifundamentls in \eref{dualmodification29}, or the diagonal ones in \eref{modification29}, and is visible already in the UV theories. This model, then, does not actually manifest any symmetry enhancement in the IR, and we present it here mostly for completeness. The index can be written as
\bes{
&1+ 2 d_1^{-2} t^{\frac{71}{75}} + 2 d_1^{2} t^{\frac{79}{75}} + d_1^{-1} \left[ \chi^{SU(2)}_{[2]}(u) +\chi^{SU(2)}_{[2]}(v)+ \chi^{SU(2)}_{[2]}(w) \right] t^{\frac{221}{150}} \\
&-2(y+y^{-1})d_1 t^{\frac{229}{150}} +d_1^{-\frac{5}{2} } \chi^{SU(2)}_{[2]}(u) \chi^{SU(2)}_{[2]}(v) \chi^{SU(2)}_{[2]}(w) t^{\frac{101}{60}} \\
& + 3 d_1^{-4} t^{\frac{142}{75}} + 2 d_1^{-2} (y+y^{-1}) t^{\frac{146}{75}} \\
& + \left[ 3-\chi^{SU(2)}_{[2]}(u) -\chi^{SU(2)}_{[2]}(v)- \chi^{SU(2)}_{[2]}(w)  \right] t^2 +\ldots~.
}
The terms at order $t^2$ indicate that the theory has the flavour symmetry $SU(2)_u \times SU(2)_v \times SU(2)_w \times U(1)_{d_1}$.

From quiver \eref{dualmodification29}, one may expect to consider instead the superpotential 
\be
W= Q_U X_L Q_D  + Q_U X_R Q_D + Q_L X_L Q_L + Q_R X_R Q_R~, 
\ee
Note that the last two cubic terms break the $SU(2)_u$ and $SU(2)_v$ flavour symmetries to $SO(2)_u$ and $SO(2)_w$ respectively.
This is actually the 4d $\CN=2$ theory with an $SU(2) \times SU(2)$ gauge group, one bifundamental hypermultiplet, and one flavour of fundamental hypermultiplet for each gauge group.  However, since each $SU(2)$ gauge group has three flavour of fundamental hypermultplets charged under it, this theory flows to a theory of two free vector multiplets (after flipping the operators $\tr(X_L^2)$ and $\tr(X_R^2)$, which fall below the unitarity bound).  The latter can be seen from the conformal anomalies: $(a,c) = (3/8,1/4)=2(3/4,1/2)$.

\subsection{A model with an enhanced $SU(2)$ symmetry}
We consider the following modification of figure 30 of \cite{Kim:2018lfo}:
\be
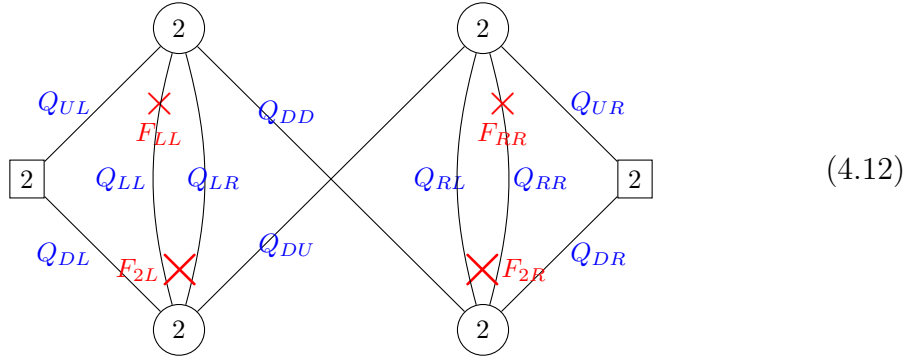
\begin{tikzpicture}[baseline]
\tikzstyle{every node}=[font=\footnotesize]
\node[draw, rectangle] (sqL) at (-4,0) {$2$};
\node[draw, circle] (cTL) at (-2,2) {$2$};
\node[draw, circle] (cTR) at (2,2) {$2$};
\node[draw, circle] (cBL) at (-2,-2) {$2$};
\node[draw, circle] (cBR) at (2,-2) {$2$};
\node[draw, rectangle] (sqR) at (4,0) {$2$};
\draw[-] (sqL) to node[left] {\blue $Q_{UL}$}(cTL);
\draw[-] (sqL) to node[left] {\blue $Q_{DL}$}(cBL);
\draw[-] (sqR) to node[right] {\blue $Q_{UR}$}(cTR);
\draw[-] (sqR) to node[right] {\blue $Q_{DR}$}(cBR);
\draw[-] (cTL) to [bend left=15]  node[xshift=-1.1cm] {\blue $Q_{LL}$} node[xshift=-0.6cm, yshift=1cm]  {\Large \red $\times$} node[xshift=-0.6cm, yshift=0.6cm]  {\red $F_{LL}$} (cBL) ;
\draw[ -] (cTL) to [bend right=15] node[xshift=0.8cm] {\blue $Q_{LR}$} node[xshift=0.35cm, yshift=-1.2cm]  {\fontsize{100}{120}\selectfont \red $\times$} node[xshift=-0.2cm, yshift=-1.2cm]  {\red $F_{2L}$} (cBL);
\draw[-] (cTL) to node[near start, xshift=0.3cm]  {\blue $Q_{DD}$}  (cBR);
\draw[-] (cBL) to node[near start, xshift=0.3cm] {\blue $Q_{DU}$} (cTR) ;
\draw[-] (cTR) to [bend left=15] node[xshift=-0.9cm] {\blue $Q_{RL}$}  node[xshift=-0.35cm, yshift=-1.2cm]  {\fontsize{100}{120}\selectfont \red $\times$} node[xshift=0.2cm, yshift=-1.2cm]  {\red $F_{2R}$}  (cBR);
\draw[-] (cTR) to [bend right=15]  node[xshift=1.1cm] {\blue $Q_{RR}$} node[xshift=0.6cm, yshift=1cm]  {\Large \red $\times$}  node[xshift=0.6cm, yshift=0.6cm]  {\red $F_{RR}$}  (cBR) ;
\end{tikzpicture}
\ee
with the superpotential
\bes{
W& = Q_{UL} Q_{LL} Q_{DL} +Q_{UR} Q_{RR} Q_{DR} + Q_{LR} Q_{DU} Q_{RL} Q_{DD} \\
& \quad + F_{LL} Q_{LL}^2 + F_{2L} Q_{LL} Q_{LR}  + F_{RR} Q_{RR}^2+ F_{2R} Q_{RL} Q_{RR}~.
}

There are three non-anomalous $U(1)$ symmetries, whose fugacities are denoted by $q_1$, $q_2$ and $x$.  Each chiral field in the quiver carries the global charges as indicated in the diagram below:
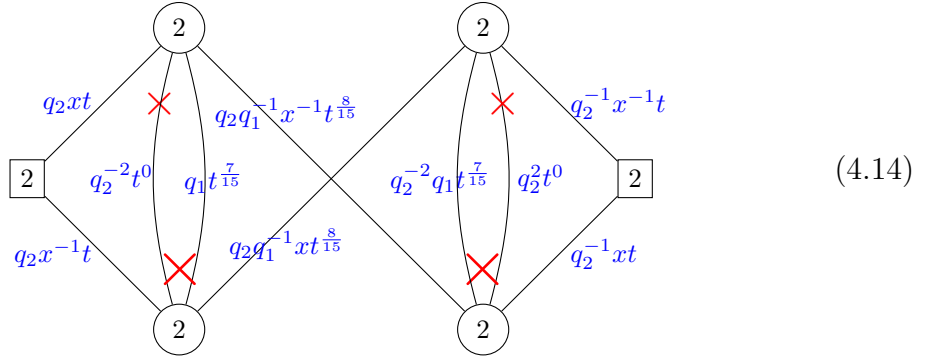
\be
\begin{tikzpicture}[baseline]
\tikzstyle{every node}=[font=\footnotesize]
\node[draw, rectangle] (sqL) at (-4,0) {$2$};
\node[draw, circle] (cTL) at (-2,2) {$2$};
\node[draw, circle] (cTR) at (2,2) {$2$};
\node[draw, circle] (cBL) at (-2,-2) {$2$};
\node[draw, circle] (cBR) at (2,-2) {$2$};
\node[draw, rectangle] (sqR) at (4,0) {$2$};
\draw[-] (sqL) to node[left] {\blue $q_2 x t $}(cTL);
\draw[-] (sqL) to node[left] {\blue $q_2 x^{-1} t $}(cBL);
\draw[-] (sqR) to node[right] {\blue $q_2^{-1} x^{-1} t $}(cTR);
\draw[-] (sqR) to node[right] {\blue $q_2^{-1}  x t$}(cBR);
\draw[-] (cTL) to [bend left=15]  node[xshift=-1.1cm] {\blue $q_2^{-2} t^0$} node[xshift=-0.6cm, yshift=1cm]  {\Large \red $\times$} (cBL) ;
\draw[ -] (cTL) to [bend right=15] node[xshift=0.8cm] {\blue $q_1 t^{\frac{7}{15}}$} node[xshift=0.35cm, yshift=-1.2cm]  {\fontsize{100}{120}\selectfont \red $\times$} (cBL);
\draw[-] (cTL) to node[near start, xshift=0.3cm]  {\blue $q_2 q_1^{-1} x^{-1} t^{\frac{8}{15}}$}  (cBR);
\draw[-] (cBL) to node[near start, xshift=0.3cm] {\blue $q_2 q_1^{-1} x t^{\frac{8}{15}}$} (cTR) ;
\draw[-] (cTR) to [bend left=15] node[xshift=-0.9cm] {\blue $q_2^{-2} q_1  t^{\frac{7}{15}}$}  node[xshift=-0.35cm, yshift=-1.2cm]  {\fontsize{100}{120}\selectfont \red $\times$}  (cBR);
\draw[-] (cTR) to [bend right=15]  node[xshift=1.1cm] {\blue $q_2^2 t^0$} node[xshift=0.6cm, yshift=1cm]  {\Large \red $\times$} (cBR) ;
\end{tikzpicture}
\ee
where the powers of the fugacity $t$ denote the \emph{approximate} superconformal $R$-charges\footnote{The exact mixing coefficients $\alpha$ such that $U(1)_R = \frac{2}{3} + \alpha U(1)_c$ for $c={q_1}, {q_2}, x$ are $\frac{1}{3} \left(\sqrt{109}-11\right) \approx -\frac{1}{5}$, $\frac{1}{3}$ and $\frac{1}{3}$, respectively.} that are used in the computation of the index below.  The conformal anomalies are
\be
(a,c) =\left(\frac{109 \sqrt{109}}{24}-\frac{369}{8},\frac{55 \sqrt{109}}{12}-\frac{93}{2} \right)~.
\ee

We claim that $U(1)_x$ gets enhanced to $SU(2)_x$ in the IR. The evidence for this is as follows.  First of all, the 't Hooft anomalies involving odd powers of $U(1)_x$ vanish, as it should be in order for $U(1)_x$ to get enhanced to $SU(2)_x$.  Moreover, one can compute the index of this theory:
\bes{ \label{indexfig30}
& 1+ (q_1^2 + q_1^2 q_2^{-4}) t^{\frac{14}{15}} + q_2^2 q_1^{-2} \chi^{SU(2)}_{[2]} (x) t^{\frac{16}{15}} - (q_1+ q_1 q_2^{-2})(y+y^{-1}) t^{\frac{22}{15}}  \\
& + (q_1^{-1}+ q_2^2 q_1^{-1}) t^{\frac{23}{15}} +(q_1^4 + q_1^4 q_2^{-8} +q_1^4 q_2^{-4}) t^{\frac{28}{15}} + (q_1^2+q_1^2 q_2^{-4})(y+y^{-1}) t^{\frac{29}{15}} \\
&+ \Big[-3-\chi^{SU(2)}_{[2]}(u) -\chi^{SU(2)}_{[2]}(v) +(q_2^2+q_2^{-2}) \chi^{SU(2)}_{[2]}(x) \\
&\qquad +q_2^{-4} \left( 1+ \chi^{SU(2)}_{[2]}(u) \right)+q_2^{4} \left( 1+ \chi^{SU(2)}_{[2]}(v) \right)\Big] t^2 +\ldots~,
}
where $u$ and $v$ are the fugacities associated with the $SU(2)$ flavour symmetry of the left and right square nodes in the quiver.   We see that the index can be written in characters of $U(1)_{q_1} \times U(1)_{q_2} \times SU(2)_u \times SU(2)_v \times SU(2)_x$, at least to the evaluated order.  Note that we do not see the negative term $-\chi^{SU(2)}_{[2]}(x)$ at order $t^2$. However, this can be accounted for by a cancellation with certain marginal operators\footnote{Let us define the combinations $(P_L)_{ab}:=(Q_{LL} Q_{LR})_{ab}$ such that the indices of the lower left gauge node are contracted and $a,b=1,2$ are the indices for the upper left gauge nodes. Similarly, we define $(P_R)_{a'b'}:=(Q_{RL} Q_{RR})_{a'b'}$ such that the indices of the upper left gauge node are contracted and $a',b'=1,2$ are the indices for the lower right gauge nodes. 
Such marginal operators in the adjoint representation of $SU(2)_x$ can be written as follows: 
\bes{
x^2 t^2: \qquad&  (P_L)_{a_1 a_2} (P_R)_{a_1'a_2'} (Q_{DU})_{b_1 b_1'} (Q_{DU})_{b_2 b_2'} \epsilon^{a_1 b_1} \epsilon^{a_2 b_2} \epsilon^{a'_1 b'_1} \epsilon^{a'_2 b'_2}~, \\
x^{0} t^2: \qquad&  (P_L)_{a_1 a_2} (P_R)_{a_1'a_2'} (Q_{DU})_{b_1 b_1'} (Q_{DD})_{b_2 b_2'} \epsilon^{a_1 b_1} \epsilon^{a_2 b_2} \epsilon^{a'_1 b'_1} \epsilon^{a'_2 b'_2}~, \\
x^{-2} t^2: \qquad&  (P_L)_{a_1 a_2} (P_R)_{a_1'a_2'} (Q_{DD})_{b_1 b_1'} (Q_{DD})_{b_2 b_2'} \epsilon^{a_1 b_1} \epsilon^{a_2 b_2} \epsilon^{a'_1 b'_1} \epsilon^{a'_2 b'_2}~.
}
Notice that these combinations do not carry fugacities $q_1$ and $q_2$, as required.}
. 

\subsection{Flowing to the $\CN=2$ $SO(4)$ gauge theory with $2$ flavours}
We consider the following modification of figure 31 of \cite{Kim:2018lfo}:
\be \label{modification31}
\begin{tikzpicture}[baseline]
\tikzstyle{every node}=[font=\footnotesize]
\node[draw, rectangle] (sqL) at (-4,0) {$2$};
\node[draw, circle] (cTL) at (-2,2) {$2$};
\node[draw, circle] (cTR) at (2,2) {$2$};
\node[draw, circle] (cBL) at (-2,-2) {$2$};
\node[draw, circle] (cBR) at (2,-2) {$2$};
\node[draw, rectangle] (sqR) at (4,0) {$2$};
\draw[-] (sqL) to node[left] {\blue $Q_{UL}$}(cTL);
\draw[-] (sqL) to node[left] {\blue $Q_{DL}$}(cBL);
\draw[-] (sqR) to node[right] {\blue $Q_{UR}$}(cTR);
\draw[-] (sqR) to node[right] {\blue $Q_{DR}$}(cBR);
\draw[-] (cTL) to [bend left=15]  node[xshift=-1.1cm] {\blue $Q_{LL}$} node[xshift=-0.6cm, yshift=1cm]  {\Large \red $\times$} node[xshift=-0.6cm, yshift=0.6cm]  {\red $F_{LL}$} (cBL) ;
\draw[ -] (cTL) to [bend right=15] node[xshift=0.8cm] {\blue $Q_{LR}$} node[xshift=0.35cm, yshift=-1.2cm]  {\fontsize{100}{120}\selectfont \red $\times$}  node[xshift=-0.2cm, yshift=-1.2cm]  {\red $F_{2L}$} (cBL);
\draw[-] (cTR) to [bend left=15] node[xshift=-0.9cm] {\blue $Q_{RL}$}  node[xshift=-0.35cm, yshift=-1.2cm]  {\fontsize{100}{120}\selectfont \red $\times$} node[xshift=0.2cm, yshift=-1.2cm]  {\red $F_{2R}$} (cBR);
\draw[-] (cTR) to [bend right=15]  node[xshift=1.1cm] {\blue $Q_{RR}$} node[xshift=0.6cm, yshift=1cm]  {\Large \red $\times$} node[xshift=0.6cm, yshift=0.6cm]  {\red $F_{RR}$} (cBR) ;
\draw[-] (cTL) to node[above] {\blue $Q_{UM}$}  (cTR) ;
\draw[-] (cBL) to node[below] {\blue $Q_{DM}$}  (cBR) ;
\draw[-] (cTL) to [bend left=15] node[near start, xshift=0.3cm]  {\blue $Q_{DD}$}  (cBR);
\draw[-] (cTL) to [bend right=15] node[near start, xshift=0.3cm]  {\blue $Q_{DU}$}  (cBR);
\end{tikzpicture}
\ee
with the superpotential
\bes{ \label{supmodification31}
W&= Q_{UL} Q_{LL} Q_{DL} +Q_{UR} Q_{RR} Q_{DR} + Q_{LR} Q_{DM} Q_{DU}+ Q_{RL} Q_{UM} Q_{DD} \\
& \quad + F_{LL} Q_{LL}^2+ F_{RR} Q_{RR}^2+F_{2L} Q_{LL} Q_{LR} +  F_{2R} Q_{RL} Q_{RR}~.
}
This theory has three non-anomalous $U(1)$ symmetries, whose fugacities are denoted by $d_1$, $d_2$ and $d_3$.  Each chiral field in the quiver carries the global charges as indicated in the diagram below:
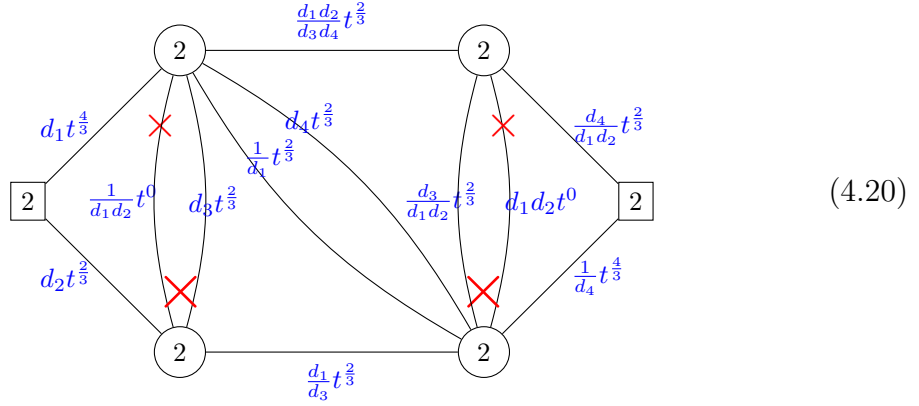
\be
\begin{tikzpicture}[baseline]
\tikzstyle{every node}=[font=\footnotesize]
\node[draw, rectangle] (sqL) at (-4,0) {$2$};
\node[draw, circle] (cTL) at (-2,2) {$2$};
\node[draw, circle] (cTR) at (2,2) {$2$};
\node[draw, circle] (cBL) at (-2,-2) {$2$};
\node[draw, circle] (cBR) at (2,-2) {$2$};
\node[draw, rectangle] (sqR) at (4,0) {$2$};
\draw[-] (sqL) to node[left] {\blue $d_1 t^{\frac{4}{3}} $}(cTL);
\draw[-] (sqL) to node[left] {\blue $d_2 t^{\frac{2}{3}} $}(cBL);
\draw[-] (sqR) to node[right] {\blue $\frac{d_4}{d_1d_2}  t^{\frac{2}{3}} $}(cTR);
\draw[-] (sqR) to node[right] {\blue $\frac{1}{d_4} t^{\frac{4}{3}}$}(cBR);
\draw[-] (cTL) to [bend left=15]  node[xshift=-1.1cm] {\blue $\frac{1}{d_1d_2} t^0$} node[xshift=-0.6cm, yshift=1cm]  {\Large \red $\times$} (cBL) ;
\draw[ -] (cTL) to [bend right=15] node[xshift=0.8cm] {\blue $d_3 t^{\frac{2}{3}}$} node[xshift=0.35cm, yshift=-1.2cm]  {\fontsize{100}{120}\selectfont \red $\times$} (cBL);
\draw[-] (cTR) to [bend left=15] node[xshift=-0.9cm] {\blue $\frac{d_3}{d_1d_2} t^{\frac{2}{3}}$}  node[xshift=-0.35cm, yshift=-1.2cm]  {\fontsize{100}{120}\selectfont \red $\times$}  (cBR);
\draw[-] (cTR) to [bend right=15]  node[xshift=1.1cm] {\blue $d_1d_2 t^0$} node[xshift=0.6cm, yshift=1cm]  {\Large \red $\times$} (cBR) ;
\draw[-] (cTL) to node[above] {\blue $\frac{d_1d_2}{d_3d_4} t^{\frac{2}{3}}$}  (cTR) ;
\draw[-] (cBL) to node[below] {\blue $\frac{d_1}{d_3} t^{\frac{2}{3}}$}  (cBR) ;
\draw[-] (cTL) to [bend left=15] node[near start, xshift=0.3cm]  {\blue $d_4 t^{\frac{2}{3}}$}  (cBR);
\draw[-] (cTL) to [bend right=15] node[near start, xshift=0.3cm]  {\blue $\frac{1}{d_1} t^{\frac{2}{3}}$}  (cBR);
\end{tikzpicture}
\ee
where the powers of the fugacity $t$ denote the \emph{exact} superconformal $R$-charges.  The conformal anomalies of this theory are
\be
(a,c) = \left( \frac{19}{12},\frac{5}{3} \right)~.
\ee
It is interesting to point out that these are coincident with those of the 4d $\CN=2$ $SO(4)$ gauge theory with $2$ flavour of hypermultiplets in the vector representation, or equivalently the $SU(2) \times SU(2)$ gauge theory with $2$ bifundamental hypermultiplets.  We will shortly describe the connection between \eref{modification31} and this $\CN=2$ theory.  The index of \eref{modification31} is
\be \label{indexmodification31}
\scalebox{0.9}{$
\begin{split}
&1+ \left(\frac{d_1^2}{d_3^2}+\frac{d_2 d_1^2}{d_3^2 d_4}+\frac{d_2^2 d_1^2}{d_3^2 d_4^2}+\frac{d_2 d_1}{d_3}+\frac{d_2^2 d_1}{d_3 d_4}+d_2^2+\frac{d_3^2}{d_1^2 d_2^2}+d_3^2+\frac{d_4^2}{d_1^2 d_2^2}+\frac{d_4}{d_2 d_3}+\frac{1}{d_3}+\frac{d_4}{d_1}  \right) t^{\frac{4}{3}}\\
&  - (y+y^{-1})\left (d_3 + \frac{d_3}{d_1 d_2} \right) t^{\frac{5}{3}}  + \Bigg[ -4-\frac{d_1^2 d_2^2}{d_3 d_4^2}-\frac{d_3 d_4^2}{d_1^2 d_2^2}-\frac{d_3 d_2}{d_1}-\frac{d_1}{d_2 d_3} \\
& \quad -\chi^{SU(2)}_{[2]}(u)  -\chi^{SU(2)}_{[2]}(v)  + \frac{1}{d_1^2 d_2^2} \left(1+\chi^{SU(2)}_{[2]} (u) \right) + {d_1^2 d_2^2} \left (1+\chi^{SU(2)}_{[2]} (v) \right) \\
& \quad + d_1 d_2 +\frac{1}{d_1 d_2} \Bigg] t^2 + \ldots~,
\end{split}$}
\ee
where $u$ and $v$ are fugacities for the $SU(2)_u$ and $SU(2)_v$ flavour symmetries denoted by the square nodes on the left and right of quiver \eref{modification31}.

In order to make a connection with the aforementioned $\CN=2$ theory, we remark that both flipping fields $F_{LL}$ and $F_{RR}$ have $R$-charge $2$, and they can be turned on in the superpotential \eref{supmodification31}, again this is assuming that there are no accidental $U(1)$ symmetries and we can trust the results of the $a$-maximisation procedure. Under the $U(1)_p = U(1)_{d_1}+U(1)_{d_2}$ symmetry (so that the fugacity $p^2=d_1d_2$), they carry charges $p^{2}$ and $p^{-2}$ respectively. Therefore there is a K\"ahler quotient implying that this combination is exactly marginal. Thus, adding $F_{LL}+F_{RR}$ in the superpotential \eref{supmodification31} amounts to moving along a one dimensional subspace of the conformal manifold.  In this subspace, $Q_{LL}$ and $Q_{RR}$ acquire a vacuum expectation value (vev).  This can be seen as follows. We have the superpotential terms $F_{LL} Q_{LL}^2+F_{RR} Q_{RR}^2 + F_{LL} + F_{RR}$, and the $F$-terms with respect to $F_{LL}$ and $F_{RR}$ force $Q^2_{LL}$ and $Q^2_{RR}$ to acquire a vev.  In other words, moving along this subspace breaks the $U(1)_p$ symmetry, and without this symmetry there is nothing that prevents $Q_{LL}$ and $Q_{RR}$ from acquiring a vev.  In either way, the vevs cause \eref{modification31} to collapse to the $\CN=2$ quiver with two $SU(2)$ gauge groups and two bifundamental hypermultiplets.  

The index of theory \eref{modification31} with the superpotential deformation $F_{LL} + F_{RR}$ in \eref{supmodification31} can be obtained from \eref{indexmodification31} by setting 
\be
d_1 = q^{1 \over 2} x~, \qquad d_2 = q^{-{1 \over 2}} x^{-1}~, \qquad d_3 = q~, \qquad d_4=q^{-{1 \over 2}} w~.
\ee
(In this parametrisation $d_1d_2=1$, and so the $U(1)_p$ symmetry defined above is broken.)  As a result, we obtain
\bes{ \label{indexSO4w2flv}
&1+ \Bigg[ 2q^2+ {\blue q^{-1} \left( \chi^{SU(2)}_{[2]}(x) + \chi^{SU(2)}_{[1]}(x) \chi^{SU(2)}_{[1]}(w) + \chi^{SU(2)}_{[2]}(w)\right)} \Bigg] t^{\frac{4}{3}} \\
& - 2q(y+y^{-1}) t^{\frac{5}{3}} + \left[ -\chi^{SU(2)}_{[2]}(x) - \chi^{SU(2)}_{[2]}(w) +2  \right] t^2 + \ldots~.
}
This is precisely equal to the index of the 4d $\CN=2$ $SU(2) \times SU(2)$ gauge theory with two bifundamental hypermultiplets, whose flavour symmetry is $USp(4)$.  Observe that the $U(1)_x$ and $U(1)_w$ symmetries of the deformed $\CN=1$ theory get enhanced to $SU(2)_x$ and $SU(2)_w$.  Indeed $SU(2)_x \times SU(2)_w$ is the subgroup of $USp(4)$ that is preserved everywhere on the conformal manifold, as can be seen from the negative terms at order $t^2$ of the index \eref{indexSO4w2flv}.  The blue terms at order $t^{\frac{4}{3}}$ correspond to the $USp(4)$ moment map operators, and the term $2q^2 t^{\frac{4}{3}}$ corresponds to the Coulomb branch operators of the two $SU(2)$ gauge groups in the $\CN=2$ theory.  The $SU(2) \times U(1)$ $R$-symmetry of the $\CN=2$ theory can indeed be decomposed into $U(1)_R \times U(1)_q$, where $U(1)_R$ is the $\CN=1$ $R$-symmetry and $U(1)_q$ commutes with $U(1)_R$.  Note that the $SU(2)_u$ and $SU(2)_v$ flavour symmetries completely decouple along the conformal manifold, as can be seen from the index \eref{indexSO4w2flv}. A way to see this is to use the fact that the only non-vanishing 't Hooft anomaly involving them is with $U(1)_p$, so once the latter is broken there is no obstruction for them to disappear in the low-energy theory.

\section{Quiver with the $E[USp(2N)]$ theory as a building block} \label{sec:EUSp2N}
Let us now consider a 4d $\CN=1$ theory whose quiver description contains the $E[USp(2N)]$ theory as a component.  The $E[USp(2N)]$ theory is a 4d $\CN=1$ SCFT with $USp(2N) \times USp(2N) \times U(1) \times U(1)$ flavour symmetry \cite{Pasquetti:2019hxf,Hwang:2020wpd}; see also appendix \ref{euspapp} for a review.   It admits a quiver description \eref{euspfields}, where only the symmetry $USp(2N) \times SU(2)^N \times U(1) \times U(1)$ is manifest.  One may use one or many copies of $E[USp(2N)]$ as a building block to construct several interesting 4d SCFTs by commonly gauging the $USp(2N)$ symmetries, including those that are not manifest in the quiver \eref{euspfields}, and couple them to matter fields\footnote{We remark that such a construction is in the same spirit of that of the 3d S-fold SCFTs in the sense that two $U(N)$ or $SU(N)$ symmetries of the $T(U(N))$ or $T(SU(N))$ theory \cite{Gaiotto:2008ak} are commonly gauged and possibly coupled to matter fields \cite{Terashima:2011qi, Gang:2015wya, Assel:2018vtq} (see also \cite{Garozzo:2018kra, Garozzo:2019hbf, Garozzo:2019ejm}).  Note that in the Lagrangian description of the $T(U(N))$ or $T(SU(N))$ theory only one $U(N)$ or $SU(N)$ symmetry is manifest, whereas the other is emergent in the IR. \label{Sfoldfootnote}}.  In \cite{Pasquetti:2019hxf}, a number of such quivers were studied in the context of compactification of the 6d rank $N$ $E$-string theory on a torus with fluxes. 

In this paper, the general strategy is as described in the preceding sections, namely we modify such quivers by lowering number of flavours (say to $N_f <8$).  The resulting quivers are expected to correspond to theories on the domain wall of the 5d $\CN=1$ $USp(2N)$ gauge theory with an antisymmetric hypermultiplet and $N_f<8$ flavours of fundamental hypermultiplets.  We also modify the superpotential and flipping fields so that the theory has interesting IR properties.  In the following, we focus on the theory that is a higher rank $USp(2N)$ generalisation of \eref{model12meq4}.  This theory turns out to have an enhanced flavour symmetry in the IR.

\subsection{A higher rank $USp(2N)$ generalization of \eref{model12meq4}}
Let us consider the following model:
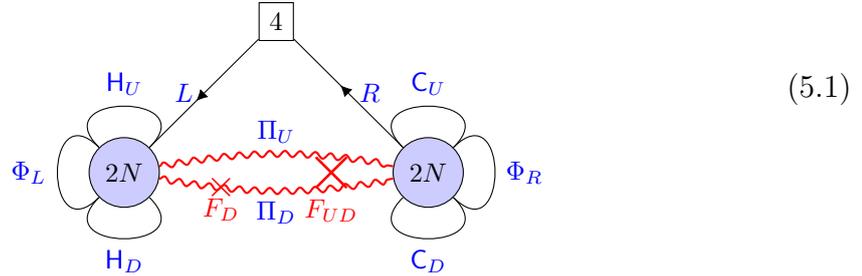
\be
\begin{tikzpicture}[baseline]
\tikzstyle{every node}=[font=\footnotesize]
\node[draw, circle, fill=blue!20] (node1) at (-2,-1) {$2N$};
\node[draw, circle, fill=blue!20] (node2) at (2,-1) {$2N$};
\node[draw, rectangle] (sqnode) at (0,1) {$4$};
 \draw[-, snake it,red, thick] (node1)  to [bend right=10] node[below] {\blue $\Pi_D$} node[below, near start] {$\red F_D$} node[near start] {\large $ \red \times$} (node2) ;
  \draw[-, snake it,red, thick] (node1) to [bend left=10] node[above] {\blue $\Pi_U$} node[below, near end, yshift=-0.4cm] {$\red F_{UD}$} node[near end, yshift=-0.2cm] {\huge $ \red \times$}  (node2);
\draw[draw=black,solid, -<-=0.5]  (node1) to node[left] {\blue $L$} (sqnode);
\draw[draw=black,solid, -<-=0.5]  (sqnode) to node[right] {\blue $R$} (node2);
\draw[black,solid] (node1) edge [out=45,in=135,loop,looseness=4] node[above]{\blue $\mathsf{H}_U$} (node1);
\draw[black,solid] (node1) edge [out=135,in=-135,loop,looseness=4] node[left]{\blue $\Phi_L$} (node1);
\draw[black,solid] (node1) edge [out=-45,in=-135,loop,looseness=4] node[below]{\blue $\mathsf{H}_D$} (node1);
\draw[black,solid] (node2) edge [out=45,in=135,loop,looseness=4] node[above]{\blue $\mathsf{C}_U$} (node2);
\draw[black,solid] (node2) edge [out=45,in=-45,loop,looseness=4] node[right]{\blue $\Phi_R$} (node2);
\draw[black,solid] (node2) edge [out=-45,in=-135,loop,looseness=4] node[below]{\blue $\mathsf{C}_D$} (node2);
\end{tikzpicture}
\ee
where we have used the notation as in appendix \ref{euspapp}.  Here two copies of $E[USp(2N)]$ are glued together by commonly gauging $USp(2N)$ symmetries from each copy, so that we have a pair of $USp(2N)$ gauge groups, denoted by blue circular nodes in the quiver.  The fields $\mathsf{H}, \mathsf{C}, \Pi$ coming from the upper (\emph{resp.} lower) copy of $E[USp(2N)]$ are labeled by  the subscripts $U$ (\emph{resp.} $D$), standing for up (\emph{resp.} down).  In the above we introduce the flipping fields $F_D$ and $F_{UD}$, as well as the chiral fields $\Gp_L$ and $\Gp_R$ in the traceless antisymmetric representation of the left and right node respectively.  The superpotential is taken to be\footnote{Contractions over $USp(2N)$ gauge indices using the antisymmetric tensor $J=\mathbb{I}_n\otimes i\,\gs_2$ and $SU(4)$ flavour indices using the Kronecker delta are understood.}:
\be
W=L\Pi_U R+\Gp_L(\mathsf{H}_U+\mathsf{H}_D)+\Gp_R(\mathsf{C}_U+\mathsf{C}_D)+F_D\Pi_D\Pi_D+F_{UD}\Pi_U\Pi_D\,,
\ee
Notice that the $F$-terms with respect to $\Phi_L$ and $\Phi_R$ have the effect of making a combination of $\mathsf{H}_U$, $\mathsf{H}_D$ and a combination of $\mathsf{C}_U$, $\mathsf{C}_D$ massive, thus leaving only one massless operator in the antisymmetric of the left gauge node and one in the antisymmetric of the right gauge node. We denote the surviving operators by $\mathsf{A}_L$ and $\mathsf{A}_R$ and we represent them in the quiver as arcs on the two nodes.  The resulting quiver is therefore
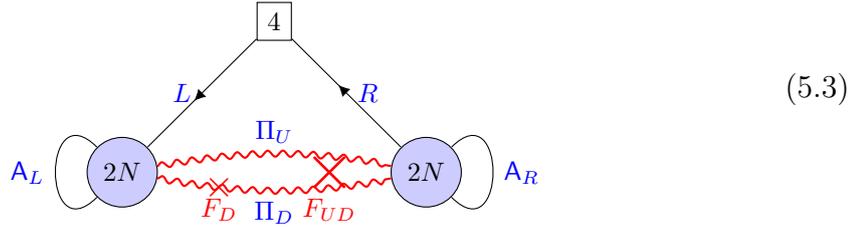
\be \label{modelwEUSp}
\begin{tikzpicture}[baseline]
\tikzstyle{every node}=[font=\footnotesize]
\node[draw, circle, fill=blue!20] (node1) at (-2,-1) {$2N$};
\node[draw, circle, fill=blue!20] (node2) at (2,-1) {$2N$};
\node[draw, rectangle] (sqnode) at (0,1) {$4$};
 \draw[-, snake it,red, thick] (node1)  to [bend right=10] node[below] {\blue $\Pi_D$} node[below, near start] {$\red F_D$} node[near start] {\large $ \red \times$} (node2) ;
  \draw[-, snake it,red, thick] (node1) to [bend left=10] node[above] {\blue $\Pi_U$} node[below, near end, yshift=-0.4cm] {$\red F_{UD}$} node[near end, yshift=-0.2cm] {\huge $ \red \times$}  (node2);
\draw[draw=black,solid, -<-=0.5]  (node1) to node[left] {\blue $L$} (sqnode);
\draw[draw=black,solid, -<-=0.5]  (sqnode) to node[right] {\blue $R$} (node2);
\draw[black,solid] (node1) edge [out=135,in=-135,loop,looseness=4] node[left]{\blue $\mathsf{A}_L$} (node1);
\draw[black,solid] (node2) edge [out=45,in=-45,loop,looseness=4] node[right]{\blue $\mathsf{A}_R$} (node2);
\end{tikzpicture}
\ee
with superpotential
\be
W=L\Pi_U R+F_D\Pi_D\Pi_D+F_{UD}\Pi_U\Pi_D\,.
\ee

The superpotential and the condition for the existence of a non-anomalous R-symmetry imply
that this theory has two non-anomalous $U(1)$ flavour symmetries, whose fugacities
we denote by $d$ and $\fug$.   The UV R-charges of the chiral fields $L$, $R$, $F_D$ and of the operators $\mathsf{A}_L$, $\mathsf{A}_R$, $\Pi_U$ and $\Pi_D$ are
\begin{eqnarray}
&&R[L]=R[R]=1-\frac{1}{2}R_d,\quad R[\mathsf{A}_L]=2-R_\fug,\quad R[\mathsf{A}_R]=R_\fug,\nn\\
&&R[\Pi_U]=R_d,\quad R[\Pi_D]=0,\quad R[F_D]=2,\quad R[F_{UD}]=2-R_d\,,
\label{rchargeseuspmodel}
\end{eqnarray}
where $R_d$ and $R_\fug$ are the mixing coefficients of the R-symmetry with the abelian global symmetries $U(1)_d$ and $U(1)_\fug$.   To relate these notations to those adopted in appendix \ref{euspapp}, we remark that the $U(1)_d$ symmetry is identified with the $U(1)_c$ symmetry of the upper copy of $E[USp(2N)]$, as can be seen by comparing $R[\Pi_U]$ in \eref{rchargeseuspmodel} with the corresponding entry in \eref{tablechargeEUSp}.  Moreover, the $U(1)_\tau$ symmetries of the upper and lower copies of $E[USp(2N)]$ are identified and are referred to as $U(1)_\tau$ in the above; this as can be seen by comparing $R[A_L]$ and $R[A_R]$ in \eref{rchargeseuspmodel} with the charges of $\mathsf{C}$ and $\mathsf{H}$ in \eref{tablechargeEUSp}.

The values of $R_d$ and $R_\fug$ that correspond to the superconformal R-charge can be determined via $a$-maximisation. For generic $N$ we find
\be \label{rchargeseuspmodelamax}
R_d=\frac{\sqrt{3 N (9 N (2 N+1)-19)+25}-9}{3 (6 N-7)},\qquad R_\fug=1\,.
\ee
For $N=1$ we recover exactly the results of section \ref{sec:enhanceSU2w4}, with the opererator $UD$ being flipped. In this case, the theory flows to the $4d$ $\CN=2$ $SU(2)$ gauge theory with four flavours.  From now on we will focus on the case $N=2$.

\subsubsection{The case of $N=2$}
We have
\be
R_d = \frac{1}{15} \left(\sqrt{451}-9\right) \approx 0.815784,\qquad R_\fug=1\,,
\label{maximizationeuspmodel}
\ee
while the conformal anomalies are
\be
(a,c)= \left(\frac{451 \sqrt{451}-5724}{1200},\frac{506 \sqrt{451}-6219}{1200} \right) \approx (0.915489, 1.17042)\,.
\ee
In order to compute the index, we approximate $R_d = \frac{4}{5}$.  Using \eref{rchargeseuspmodel}, we summarise the charges of each chiral field as follows:
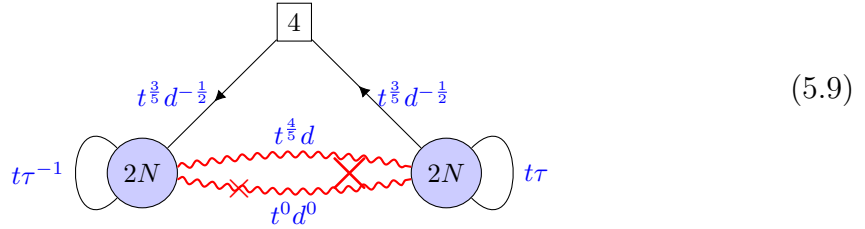
\be \label{chargeblock}
\begin{tikzpicture}[baseline]
\tikzstyle{every node}=[font=\footnotesize]
\node[draw, circle, fill=blue!20] (node1) at (-2,-1) {$2N$};
\node[draw, circle, fill=blue!20] (node2) at (2,-1) {$2N$};
\node[draw, rectangle] (sqnode) at (0,1) {$4$};
 \draw[-, snake it,red, thick] (node1)  to [bend right=10] node[below] {\blue $t^0 d^0$} node[below, near start] {} node[near start] {\large $ \red \times$} (node2) ;
  \draw[-, snake it,red, thick] (node1) to [bend left=10] node[above] {\blue $t^{\frac{4}{5}} d$} node[below, near end, yshift=-0.4cm] {} node[near end, yshift=-0.2cm] {\huge $ \red \times$}  (node2);
\draw[draw=black,solid, -<-=0.5]  (node1) to node[left] {\blue $t^{\frac{3}{5}} d^{-\frac{1}{2}}$} (sqnode);
\draw[draw=black,solid, -<-=0.5]  (sqnode) to node[right] {\blue $t^{\frac{3}{5}} d^{-\frac{1}{2}}$} (node2);
\draw[black,solid] (node1) edge [out=135,in=-135,loop,looseness=4] node[left]{\blue $t \tau^{-1}$} (node1);
\draw[black,solid] (node2) edge [out=45,in=-45,loop,looseness=4] node[right]{\blue $t \tau$} (node2);
\end{tikzpicture}
\ee
where the powers of $t$ denote the approximate $R$-charges.  Using the charge assignment as in \eref{chargeblock}, we find that the index is
\bes{
&1+d^{-1}\left[ \chi^{SU(4)}_{[1,0,1]} (\vec u)+ 2\chi^{SU(4)}_{[0,1,0]} (\vec u) +2\right] t^{\frac{6}{5}} + d^{-2} (\tau + \tau^{-1}) \,t^{\frac{7}{5}}+2d^2t^{\frac{8}{5}} \\
& \quad -d\left(y+y^{-1}\right)t^{\frac{9}{5}} - t^2+ d^{-1}\Big[(\tau+\tau^{-1}) \left( \chi^{SU(4)}_{[1,0,1]} (\vec u)+ 2\chi^{SU(4)}_{[0,1,0]} (\vec u) \right) \\
& \quad +(y+y^{-1}) \left( \chi^{SU(4)}_{[1,0,1]}(\vec u)+ 2\chi^{SU(4)}_{[0,1,0]} (\vec u) +2 \right)\Big]t^{\frac{11}{5}} +\ldots
}
where $\vec u = (u_1,u_2,u_3)$ denotes the $SU(4)$ fugacities.  Recalling the following branching rule of the adjoint representation of $SO(8)$ to $SU(4) \times U(1)$:
\bes{
[0,1,0,0] \quad &\to \quad  [1,0,1]_0 + [0,1,0]_{+2} + [0,1,0]_{-2}+[0,0,0]_0\\
{\bf 28} \quad &\to \quad {\bf 15}_0\oplus{\bf 6}_{+2}\oplus {\bf 6}_{-2} \oplus{\bf 1}_0~,
}
we claim that the $SU(4)$ flavour symmetry in the description \eref{chargeblock} gets enhanced to $SO(8)$ in the IR.  Note that the aforementioned $U(1)$, which is a commutant of $SU(4)$ in $SO(8)$, is not manifest in the description \eref{chargeblock}; it is in fact emergent in the IR and combines with $SU(4)$ to become $SO(8)$.  Moreover, we claim that the $U(1)_\tau$ gets enhanced to $SU(2)_\tau$.  Indeed, the above index can be rewritten as
\be\label{characteradjointso8}
\scalebox{0.95}{$
\begin{split}
&1+d^{-1}\left[ \chi^{SO(8)}_{[0,1,0,0]} (\vec x) +1 \right] t^{\frac{6}{5}}+ d^{-2}\chi^{SU(2)}_{[1]} (\fug)\,t^{\frac{7}{5}}+2d^2t^{\frac{8}{5}}-d\left(y+y^{-1}\right)t^{\frac{9}{5}} \\
& \quad - t^2+d^{-1}\left[\chi^{SU(2)}_{[1]} (\fug)(\chi^{SO(8)}_{[0,1,0,0]} (\vec x)-1)+(y+y^{-1})(\chi^{SO(8)}_{[0,1,0,0]} (\vec x)+1)\right]t^{\frac{11}{5}} +\ldots~.
\end{split}
$}
\ee
Let us now discuss the symmetry enhancement in further detail.

We first consider the enhancement of $SU(4)$ to $SO(8)$. Note that such enhancement also occurs in the $N=1$ case, as discussed in section \ref{sec:enhanceSU2w4}, where the theory flows to 4d $\CN=2$ $SU(2)$ gauge theory with four flavours, whose flavour symmetry is $SO(8)$.   First of all, we notice that the index rearranges into characters of $SO(8)$.  For example, at order $t^{\frac{6}{5}}$, we have the terms $d^{-1}\left( \chi^{SO(8)}_{[0,1,0,0]} (\vec x)+1 \right)$, which come from the following operators in the following representations of $SU(4) \times U(1)_d$:
\be \label{t65eusp}
\begin{array}{ll}
d^{-1}\chi^{SU(4)}_{[0,1,0]} (\vec u)t^{\frac{6}{5}}:  \qquad \qquad & (LL)^{ij}=L^i_aL^j_bJ^{a\,b} \\
d^{-1}\chi^{SU(4)}_{[0,1,0]} (\vec u)t^{\frac{6}{5}}: \qquad  \qquad& (RR)_{ij}=R_{i\,a'}R_{j\,b'}J^{a'\,b'} \\
 d^{-1}(\chi^{SU(4)}_{[1,0,1]} (\vec u)+1)t^{\frac{6}{5}}: \qquad  \qquad& (L\Pi_DR)^i{}_j=L^i_a\Pi_{D,b\,a'}R_{j\,b'}J^{a\,b}J^{a'\,b'} \\
d^{-1}t^{\frac{6}{5}}: \qquad  \qquad& F_{UD}
\end{array}
\ee
where $a,b=1,\ldots,4$ and $a',b'=1,\ldots,4$ are the indices for the left and right $USp(4)$ gauge nodes respectively.
Moreover, in order for the enhancement to hold we should see the contribution of the conserved current in the adjoint representation of $SO(8)$ contributing with a minus sign at order $t^2$, while from the index we only see $-t^2$ which we interpret as the contribution of the conserved current for the $U(1)_d$ symmetry. Nevertheless, the absence of this contribution to the index might be attributed to cancellations with some marginal operators in the adjoint representation of $SO(8)$ and uncharged under $U(1)_d$\footnote{Let us define the combination
\be
P_{ab'} =  \Pi_{U,a_1\,a_1'} \Pi_{D,a_2\,b'} \Pi_{D,a\,a_3'} J^{a_1\,a_2} J^{a_1'\,a_3'} ~.
\ee
Such marginal operators and their fugacities are as follows:
\be \label{t2eusp}
\begin{array}{ll}
 \chi^{SU(4)}_{[0,1,0]} (\vec u)t^2: \qquad \quad &(L\Pi_U\Pi_D \Pi_D\Pi_D L)^{ij}=L^i_{b_1} P_{a_1 a'_1} \Pi_{D,a_2\,a_2'} L^j_{b_2} J^{a_1\,b_1} J^{a_2\,b_2} J^{a_1' \, a_2'}  \\
\chi^{SU(4)}_{[0,1,0]} (\vec u)t^2: \qquad  \quad & (R\Pi_U\Pi_D \Pi_D\Pi_DR)_{ij}=R_{i\,b_1'} P_{a_1 a'_1} \Pi_{D,a_2\,a_2'}  R_{j\,b_2'}J^{a_1\,a_2} J^{a_1'\,b_1'} J^{a_2'\,b_2'} \\
 \chi^{SU(4)}_{[1,0,1]} (\vec u)t^2: \qquad \quad& (L\Pi_U\Pi_D\Pi_D R)^i{}_j=L^i_{b_1} P_{a_1 a_1'}R_{j\,b_1'}J^{a_1\,b_1}J^{a_1'\,b_1'}  \\
 t^2: \qquad \quad &  F_D.
\end{array} 
\ee
Notice that these gauge invariant combinations are neutral under $U(1)_d$, as required.
}.

Regarding the enhancement from $U(1)_\fug$  to $SU(2)_\tau$, we again notice that the index rearranges into characters of $SU(2)_\tau$. In particular, at order $t^{\frac{7}{5}}$ we see an operator in the fundamental representation of $SU(2)_\tau$, which is made of the two following gauge invariant operators of the upper $E[USp(2N)]$ block (see appendix \ref{euspapp}):
\begin{eqnarray}
d^{-2}\fug\, t^{\frac{7}{5}}: & \qquad b_1^{(U)}\\
d^{-2}\fug^{-1} t^{\frac{7}{5}}: & \qquad \mathsf{M}_1^{(U)} ~.
\end{eqnarray}
where the superscript $(U)$ is there to emphasize that these are operators coming from the upper $E[USp(2N)]$ theory.
Note also that the 't Hooft anomalies involving odd powers of $U(1)_\tau$ vanish. This is indeed a necessary condition for the enhancement to $SU(2)_\tau$. Finally, we again note that we do not observe the conserved currents for this $SU(2)_\tau$ in the $t^2$ order in the index. This again might be explained by a cancellation with some marginal operators. For instance, there are the marginal operators, with their index contributions:
\bes{ \label{SU2marginal}
 \tau^{-2}t^2: \qquad \quad& \mathsf{A}_L\Pi_D\Pi_D\mathsf{A}_L=\mathsf{A}_{L,a\,b}\mathsf{A}_{L,c\,d}\Pi_{D,e\,a'}\Pi_{D,f\,b'}J^{a\,e}J^{c\,f}J^{b\,d}J^{a'\,b'} \\
 \tau^{2}t^2: \qquad \quad &  \mathsf{A}_R\Pi_D\Pi_D\mathsf{A}_R=\mathsf{A}_{R,a'\,b'}\mathsf{A}_{R,c'\,d'}\Pi_{D,a\,e'}\Pi_{D,b\,f'}J^{a'\,e'}J^{c'\,f'}J^{b'\,d'}J^{a\,b} \\
 \tau^{0} t^2: \qquad \quad & \mathsf{A}_L\Pi_D \Pi_D\mathsf{A}_R=\mathsf{A}_{L,a\,b}\mathsf{A}_{R,a'\,b'}\Pi_{D,c\,c'}\Pi_{D,d\,d'}J^{a\,c}J^{a'\,c'}J^{b\,d}J^{b'\,d'}~.
}
These could cancel the contribution of the conserved current in the adjoint representation of $SU(2)_{\tau}$.

\subsubsection{General $N$}
Let us briefly comment on the case of a general value of $N$. 

We claim that the $U(1)_\tau$ gets enhanced to $SU(2)_\tau$ in the IR.  The reasons are as follows. Notice that the vanishing of the 't Hooft anomalies with odd powers of $U(1)_\tau$ holds for any $N$, and so the necessary condition for such enhancement is satisfied.  Moreover, from \eqref{rchargeseuspmodel} and \eref{rchargeseuspmodelamax}, we have $R[\Pi_D]=0$ and $R_\fug=1$ for any $N$, and so we will have the same set of marginal operators \eqref{SU2marginal} in the triplet of $SU(2)_\tau$ for general $N$. Finally, $E[USp(2N)]$ enjoys a self-duality (see appendix \ref{euspapp}) that acts on the $\fug$ fugacity of the index as $\fug\to pq/\fug = t^2/\tau$, which implies that $\tau$ will appear in the index of our model with characters of $SU(2)_\tau$. All these facts suggest that the enhancement of $U(1)_\fug$ to $SU(2)_\tau$ may also occur for higher $N$.

Regarding the enhancement of $SU(4)$ to $SO(8)$, we do not have crystal clear evidence for it taking place for $N \geq 3$.  This is partly because it is very cumbersome to compute the index for $E[USp(2N)]$ for $N \geq 3$ as a power series in $t$ to a satisfactory order.   Nevertheless, one can still see some signals of the $SO(8)$ symmetry.  For example, the relevant operators \eqref{t65eusp} still combine into the adjoint representation of $SO(8)$ plus a singlet. The marginal operators \eqref{t2eusp} also combine into the adjoint representation of $SO(8)$, and their contribution in the index still cancels that of the possible $SO(8)$ conserved current for any $N$. Indeed, these signals are due to the fact that we gave $R[L]=R[R]=2-R[\Pi_U]$ and $R[\Pi_D]=0$ for generic $N$; see \eqref{rchargeseuspmodel}.


\section{A model with an enhanced $SU(9)$ symmetry}  \label{sec:SU9enhancement}
In this section, we consider a quiver theory with a $USp(4) \times SU(3)$ gauge group that is a variation of figure 4(b) of \cite{Kim:2018bpg} and figure 6 of \cite{Zafrir:2018hkr}, associated with the $(D_5,D_5)$ conformal matter on a torus with flux $\frac{1}{2}$.  The modification is such that the gauge anomalies are cancelled.  In particular, we study the following model:
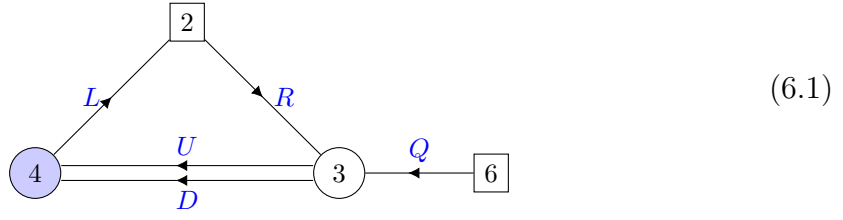
\be
\begin{tikzpicture}[baseline]
\tikzstyle{every node}=[font=\footnotesize]
\node[draw, circle, fill=blue!20] (node1) at (-2,-1) {$4$};
\node[draw, circle] (node2) at (2,-1) {$3$};
\node[draw, rectangle] (sqnode) at (0,1) {$2$};
\node[draw, rectangle] (sqnode2) at (4,-1) {$6$};
\draw[transform canvas={yshift=-2.8pt}, -<-=0.5] (node1) to  node[below] {\blue $D$}  (node2) ;
  \draw[transform canvas={yshift=2.8pt}, -<-=0.5] (node1) to node[above] {\blue $U$}  (node2);
\draw[draw=black,solid, ->-=0.5]  (node1) to node[left] {\blue $L$} (sqnode);
\draw[draw=black,solid, ->-=0.5]  (sqnode) to node[right] {\blue $R$} (node2);
\draw[draw=black,solid, -<-=0.5]  (node2) to node[above] {\blue $Q$} (sqnode2);
\end{tikzpicture}
\ee
where the blue circular node with the label 4 denotes the $USp(4)$ gauge group, and the white circular node with the label 3 denotes the $SU(3)$ gauge group. 
Let us first focus on the zero superpotential case:
\be
W=0
\ee
The condition for the non-anomalous $R$-symmetry implies that the $R$-charges of the chiral fields can be written as
\be
R[(U,D,L,R, Q)]  = \left( x+\frac{2}{3},y+\frac{2}{3},-\frac{3 x}{2}-\frac{3 y}{2}-1,z+\frac{2}{3},-\frac{2 x}{3}-\frac{2 y}{3}-\frac{z}{3}+\frac{5}{9} \right)
\ee
$a$-maximisation fixes $(x,y,z)$ to be $\left( \frac{\sqrt{10}}{9}-\frac{3}{4},\frac{\sqrt{10}}{9}-\frac{3}{4}, \frac{1}{9} \left(6-\sqrt{10}\right) \right)$ and so
\bes{
R[(U,D,L,R, Q)]  &= \left( \frac{\sqrt{10}}{9}-\frac{1}{12},\frac{\sqrt{10}}{9}-\frac{1}{12},\frac{5}{4}-\frac{\sqrt{10}}{3},\frac{4}{3}-\frac{\sqrt{10}}{9},\frac{4}{3}-\frac{\sqrt{10}}{9} \right) \\
&\approx (0.268,0.268,0.196,0.982,0.982)~.
}
Observe that the gauge invariant combination $LL$ has $R$-charge $0.392$, falling below the unitarity bound. We therefore introduce the flipping field $F_L$ and add the superpotential term $F_L(LL)$.
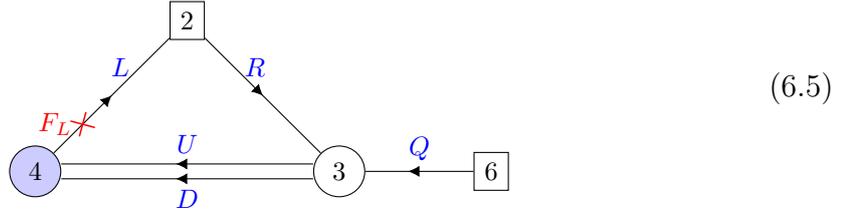
\be \label{C2A2}
\begin{tikzpicture}[baseline]
\tikzstyle{every node}=[font=\footnotesize]
\node[draw, circle, fill=blue!20] (node1) at (-2,-1) {$4$};
\node[draw, circle] (node2) at (2,-1) {$3$};
\node[draw, rectangle] (sqnode) at (0,1) {$2$};
\node[draw, rectangle] (sqnode2) at (4,-1) {$6$};
\draw[transform canvas={yshift=-2.8pt}, -<-=0.5] (node1) to  node[below] {\blue $D$}  (node2) ;
  \draw[transform canvas={yshift=2.8pt}, -<-=0.5] (node1) to node[above] {\blue $U$}  (node2);
\draw[draw=black,solid, ->-=0.5]  (node1) to node[near start,left] {$\red F_L$} node[near start, rotate=30] {\large $ \red \times$}  node[left, near end] {\blue $L$} (sqnode);
\draw[draw=black,solid, ->-=0.5]  (sqnode) to node[right, near start] {\blue $R$} (node2);
\draw[draw=black,solid, -<-=0.5]  (node2) to node[above] {\blue $Q$} (sqnode2);
\end{tikzpicture}
\ee
with
\be
W = F_L(LL)~.
\ee
$a$-maximisation fixes $(x,y,z)$ to be $\left( -\frac{83}{216},  -\frac{83}{216},  \frac{65}{216} \right)$ and so
\bes{
R[(U,D,L,R, Q, F_L)]  &= \left( \frac{61}{216},\frac{61}{216},\frac{11}{72},\frac{209}{216},\frac{209}{216},\frac{61}{36} \right) \\
&\approx (0.282,0.282,0.153,0.968,0.968,1.694)~.
}
The conformal anomalies are
\be \label{acpreoriginal}
(a,c) = \left( \frac{1909}{1024},\frac{6895}{3072} \right) \approx  (1.864,2.244)~.
\ee

\paragraph{Adding the superpotential term $ULR$.} Let us deform the theory by turning on the relevant deformation $ULR$, whose $R$-charge is $\frac{101}{72} \approx 1.403$, in the superpotential so that
\bes{ \label{supC2A2}
W &= ULR + F_L(LL)~.
}
If we {\it assume} that there is no accidental symmetry, $a$-maximisation gives $(x,y)\approx(-0.428,0.034)$ and so
\be \label{RC2A2}
R[(U,D,L,R, Q, F_L)] \approx (0.239,~0.371,~0.085,~1.676,~0.701,~1.830)~.
\ee
These lead to the conformal anomalies
\be \label{acoriginal}
(a,c) = (2.167,~2.573)~.
\ee
We will see below that there is, in fact, an accidental symmetry.  This renders the $R$-charges \eref{RC2A2} obtained using $a$-maximisation {\it unreliable}\footnote{One hint that there is something wrong with these results can already be seen as \eref{acoriginal} is larger than \eref{acpreoriginal}, in contradiction with the $a$-theorem.}.  To understand this point, it is more transparent to consider the Intriligator--Pouliot dual \cite{Intriligator:1995ne} of \eref{C2A2}.

\subsection{Intriligator--Pouliot dual of theory \eref{C2A2}}
We apply the Intriligator--Pouliot duality \cite{Intriligator:1995ne} to the $USp(4)$ gauge group of \eref{C2A2}.  Recall that, under this duality, the $USp(4)$ SQCD with 8 fundamentals is a Wess--Zumino model with 28 chiral multiplets, represented by an $8 \times 8$ antisymmetric matrix $M$, with the quartic superpotential $W= \mathrm{Pf} \, M$.  The dual of model \eref{C2A2} can be written as
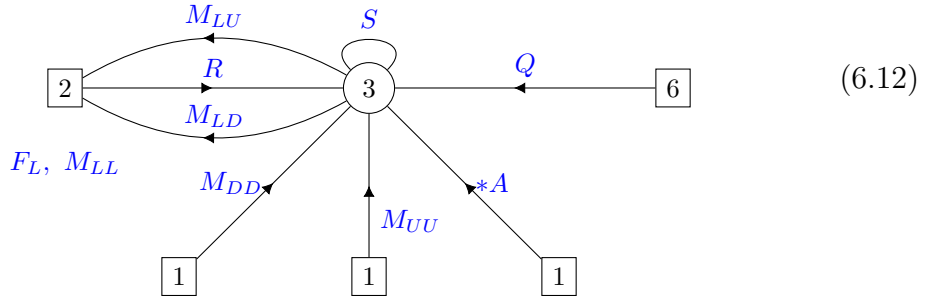
\be \label{dualC2A2}
\begin{tikzpicture}[baseline]
\tikzstyle{every node}=[font=\footnotesize]
\node[draw, rectangle] (sqnodeL) at (-4,0) {$2$};
\node[draw=none] at (-4,-1) {\blue $F_L, ~M_{LL}$};
\node[draw, rectangle] (sqnode1) at (0,-2.5) {$1$};
\node[draw, rectangle] (sqnode2) at (-2.5,-2.5) {$1$};
\node[draw, rectangle] (sqnode3) at (2.5,-2.5) {$1$};
\node[draw, circle] (node) at (0,0) {$3$};
\node[draw, rectangle] (sqnodeR) at (4,0) {$6$};
\draw[transform canvas={yshift=0pt}, ->- =0.5] (sqnode1) to  node[right, near start]{\blue $M_{UU}$} (node);
\draw[transform canvas={yshift=0pt}, -<-=0.5] (sqnodeL) to [bend left=30] node[above] {\blue $M_{LU}$}(node);
\draw[transform canvas={yshift=0pt}, -<-=0.5] (sqnodeL) to [bend right=30] node[above] {\blue $M_{LD}$}(node);
\draw[transform canvas={yshift=0pt}, ->- =0.5 ]  (sqnode2) to  node[left]{\blue $M_{DD}$} (node);
\draw[transform canvas={yshift=0pt}, ->- =0.5 ]  (sqnode3) to  node[right]{\blue $*A$} (node);
\draw[transform canvas={yshift=0pt}, -<- = 0.5]  (node) to node[above]{\blue $Q$}  (sqnodeR);
\draw[transform canvas={yshift=0pt}, ->- = 0.5]  (sqnodeL) to node[above]{\blue $R$}  (node);
\draw[black,solid] (node) edge [out=45,in=135,loop,looseness=4] node[above]{\blue $S$}   (node);
\end{tikzpicture}
\ee
where $M_{X}$ denotes the components of $M$ dual to the bilinear $X$ in \eref{C2A2}. The combination $UD$ (with the $USp(4)$ gauge indices contracted) can be decomposed into a rank-two symmetric field $S$ and a rank-two antisymmetric field $A$ under $SU(3)$.  Note that the latter can also be regarded as a chiral field $*A$ in the antifundamental representation of $SU(3)$.  The 28 components of $M$ therefore split as follows: $6$ of $M_{LU}$, $6$ of $M_{LD}$, $3$ of $M_{UU}$, $3$ of $M_{DD}$, $1$ of $M_{LL}$, $6$ of $S$, and $3$ of $A$. 
The superpotential of this theory can be determined by putting all of the possible gauge and flavour invariants that map to the combinations of the fields in \eref{C2A2} with $R$-charge $2$ and $U(1)_x$, $U(1)_y$ charge $0$:
\bes{
W &=  ~M_{LU}  (S^2 M_{LD} + AS M_{LD}+ A^2 M_{LD}+M_{UU}M_{DD}M_{LD}) \\
&~+M_{LU}^2(M_{DD} S+M_{DD} A) \\ 
&~+M_{LL} (S^3 + A S^2+ A^2S + A^3 + M_{DD}M_{UU} S+ M_{DD}M_{UU} A)\\
&~+ M_{UU} A M_{LD}^2 + M_{UU} S M_{LD}^2 \\
&~+M_{LL} F_L ~.
}

Let us now consider the dual of the theory with superpotential \eref{supC2A2}.  In the latter, the superpotential term $ULR=M_{LU}R$ implies that the fields $R$ and $M_{LU}$ acquire a mass and so can be integrated out\footnote{Note that the matter content in \eref{dualC2A2} renders the $SU(3)$ gauge group to be IR free, and so, without the superpotential \eref{supC2A2}, the theory has trivial dynamics in the IR. This implies that the results in \eref{acpreoriginal} are inaccurate.}. 
%
The resulting quiver is then
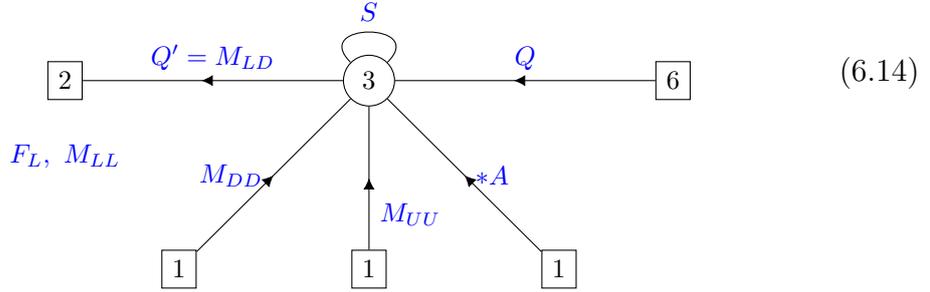
\be \label{dualC2A2a}
\scalebox{1}{
\begin{tikzpicture}[baseline]
\tikzstyle{every node}=[font=\footnotesize]
\node[draw, rectangle] (sqnodeL) at (-4,0) {$2$};
\node[draw, rectangle] (sqnode1) at (0,-2.5) {$1$};
\node[draw, rectangle] (sqnode2) at (-2.5,-2.5) {$1$};
\node[draw, rectangle] (sqnode3) at (2.5,-2.5) {$1$};
\node[draw, circle] (node) at (0,0) {$3$};
\node[draw, rectangle] (sqnodeR) at (4,0) {$6$};
\node[draw=none] at (-4,-1) {\blue $F_L, ~M_{LL}$};
\draw[transform canvas={yshift=0pt}, ->- =0.5] (sqnode1) to  node[right, near start]{\blue $M_{UU}$} (node);
\draw[transform canvas={yshift=0pt}, -<-=0.5] (sqnodeL) to node[above] {\blue $Q' = M_{LD}$}(node);
\draw[transform canvas={yshift=0pt}, ->- =0.5 ]  (sqnode2) to  node[left]{\blue $M_{DD}$} (node);
\draw[transform canvas={yshift=0pt}, ->- =0.5 ]  (sqnode3) to  node[right]{\blue $*A$} (node);
\draw[transform canvas={yshift=0pt}, -<- = 0.5]  (node) to node[above]{\blue $Q$}  (sqnodeR);
\draw[black,solid] (node) edge [out=45,in=135,loop,looseness=4] node[above]{\blue $S$}   (node);
\end{tikzpicture}
}
\ee
with the superpotential
\bes{ \label{supdualC2A2a}
W_{\eref{dualC2A2a}} = & M_{LL} (S^3 + A S^2+ A^2S + A^3 + M_{DD}M_{UU} S+ M_{DD}M_{UU} A)\\
&~+ M_{UU} A M_{LD}^2 + M_{UU} S M_{LD}^2 \\
&~+M_{LL} F_L ~.
}
Indeed, we see that dualising the $USp(4)$ gauge group in the original theory brings about quartic superpotential terms, which correspond to irrelevant operators with respect to the UV fixed point. These indeed lead to an accidental symmetry.

The superpotential and the condition for non-anomalous $R$-symmetry give
\bes{ \label{RdualC2A2a}
&R[(Q, M_{UU}, M_{DD}, A, M_{LD}, S, M_{LL}, F_L)]\\ 
&= \Big(-\frac{5 x}{12}-\frac{7 y}{12}+\frac{2}{3},~x+\frac{2}{3},~y+\frac{2}{3},~\frac{x}{2}+\frac{y}{2}+\frac{2}{3}, \\
& \qquad \qquad \qquad -\frac{3 x}{4}-\frac{y}{4}+\frac{1}{3},~\frac{x}{2}+\frac{y}{2}+\frac{2}{3},~-\frac{3 x}{2}-\frac{3 y}{2},~\frac{3 x}{2}+\frac{3 y}{2}+2 \Big)~.
}
If we were to proceed with $a$-maximisation, we would obtain $(x,y) \approx (-0.189,~ 0.076)$, and so
\bes{ \label{valuesRdualC2A2a}
&R[(Q, M_{UU}, M_{DD}, A, M_{LD}, S, M_{LL}, F_L)]\\ 
&\approx (0.701,~0.478,~0.743,~0.610,~0.456,~0.610,~0.169,~1.830)~.
}
With these values of the $R$-charges, we would obtain the conformal anomalies
\be 
(a,c) = (2.167,~2.573)~,
\ee
in agreement with \eref{acoriginal}.  However, due to the accidental symmetry, the $R$-charges presented in \eref{valuesRdualC2A2a} are {\it unreliable}\footnote{Another piece of evidence that something goes wrong is the supersymmetric index.  Computing the index of theory \eref{dualC2A2a} with the $R$-charges \eref{valuesRdualC2A2a} and expanding it as a power series in $t = (pq)^{\frac{1}{2}}$, we obtain negative terms at the power lower than $t^2$. This is in contradiction with the superconformal symmetry.}.

\subsubsection*{An enhanced $SU(9)$ flavour symmetry} 
We claim that the theory \eref{dualC2A2a} with superpotential \eref{supdualC2A2a} flows to a superconformal field theory with a global symmetry $SU(9) \times SU(2) \times U(1)^2$, where the $U(1)^3 \times SU(6)$ flavour symmetry manifest as rectangular nodes\footnote{In fact, one of such $U(1)$ symmetries is broken by quartic superpotential terms.  However, since the latter are irrelevant, we gain this factor of $U(1)$ back in the IR.} in the quiver \eref{dualC2A2a} gets enhanced to $SU(9)$ in the IR.  Let us explain this as follows.

Let us consider \eref{dualC2A2a}, without the singlets $F_L$ and $M_{LL}$, and with zero superpotential.  We can combine $M_{DD}$, $M_{UU}$, $*A$ and $Q$, which transform in the antifundamental representation of the $SU(3)$ gauge group, into a the chiral field $F$ in the following theory:
\be \label{dualC2A2b}
\begin{tikzpicture}[baseline]
\tikzstyle{every node}=[font=\footnotesize]
\node[draw, rectangle] (sqnodeL) at (-4,0) {$2$};
\node[draw, circle] (node) at (0,0) {$SU(3)$};
\node[draw, rectangle] (sqnodeR) at (4,0) {$9$};
\draw[transform canvas={yshift=0pt}, -<-=0.5] (sqnodeL) to node[above] {\blue $Q'$}(node);
\draw[transform canvas={yshift=0pt}, -<- = 0.5]  (node) to node[above]{\blue $F$}  (sqnodeR);
\draw[black,solid] (node) edge [out=45,in=135,loop,looseness=4] node[above]{\blue $S$}   (node);
\end{tikzpicture}
\ee
with zero superpotential.  The condition for the non-anomalous $R$-symmetry fixes the $R$-charges of the chiral fields to be of the form:
\bes{ \label{RQQS}
R[(F,Q',S)] =  \left( -\frac{2 \alpha}{9}-\frac{5 \beta}{9}+\frac{16}{27},~\alpha+\frac{2}{3},~ \beta+\frac{2}{3} \right)~.
}
There are two non-anomalous $U(1)$ symmetries, denoted by $U(1)_\alpha$ and $U(1)_\beta$, under which the charges of the chiral fields are given by the corresponding coefficients of $\alpha$ and $\beta$ in the above equation.

$a$-maximisation fixes $(\alpha,\beta)$ to be 
\be
(\alpha,\beta) = \left( \frac{1}{117} \left(5 \sqrt{321}-93\right), \frac{1}{117} \left(189-11 \sqrt{321}\right) \right) \approx (-0.029,-0.069)~,
\ee
and so the superconformal $R$-charges are
\bes{
R[F] = R[Q'] &= \frac{5}{117} \left(\sqrt{321}-3\right) \approx 0.637~, \\
 R[S] &= \frac{1}{117} \left(267-11 \sqrt{321}\right) \approx 0.598~.
}
The conformal anomalies are
\be
(a,c) = \left( \frac{1177 \sqrt{321}-14880}{2704},\frac{517 \sqrt{321}-5620}{1352} \right) \approx (2.296, 2.694)~.
\ee
To compute the index of \eref{dualC2A2b}, we choose the values of the $R$-charges of the chiral fields to be close to the superconformal ones.  For convenience, we take $(\alpha,\beta)$ in \eref{RQQS} to be $(-\frac{3}{100}, -\frac{7}{100})$.  We also denote the fugacities of $U(1)_\alpha$ and $U(1)_\beta$ as $\alpha$ and $\beta$.  With these values and notations, we obtain the index to be
\be \label{indexSU9}
\scalebox{0.9}{$
\begin{split}
& 1+ \underbrace{\alpha^{\frac{7}{9}} \beta^{-\frac{5}{9}} \chi^{SU(9)}_{\vec 9} (\vec s)  \chi^{SU(2)}_{\vec 2} (\vec v) t^{\frac{1721}{1350}}}_{FQ'} +\underbrace{\beta^3 t^{\frac{179}{100}}}_{S^3} +\underbrace{ \alpha^{-\frac{4}{9}} \beta^{-\frac{1}{9}} \chi^{SU(9)}_{\vec{45}} (\vec s) t^{\frac{5057}{2700}} }_{SFF} +\underbrace{ \alpha^{-\frac{2}{3}} \beta^{-\frac{5}{3}} \chi^{SU(9)}_{\vec{84}} (\vec s) t^{\frac{1723}{900}} }_{FFF}  \\
& +\left[ -\chi^{SU(9)}_{\vec {80}} (\vec s) -  \chi^{SU(2)}_{\vec 3} (\vec v) -2 \right]t^2 \\
& - (y+y^{-1}) \alpha \beta  \chi^{SU(2)}_{\vec 2} (\vec v) + (y+y^{-1}) \alpha^{\frac{7}{9}} \beta^{-\frac{5}{9}}  \chi^{SU(9)}_{\vec 9} (\vec s) \chi^{SU(2)}_{\vec 2} (\vec v)    \\
& + \underbrace{\alpha^2 \beta^2 \chi^{SU(2)}_{\vec 3} (\vec v) t^{\frac{37}{15}}}_{Q'^2S^2} +\underbrace{\alpha^{\frac{16}{9}} \beta^{\frac{4}{9}} \chi^{SU(9)}_{\vec 9} (\vec s) t^{\frac{1693}{675}}}_{F S^2 Q'^2}+\underbrace{\alpha^{\frac{14}{9}} \beta^{-\frac{10}{9}}[ \chi^{SU(9)}_{\vec{45}} (\vec s) \chi^{SU(2)}_{\vec 3}(\vec v) +  \chi^{SU(9)}_{\vec{36}} (\vec s)  ]t^{\frac{1721}{675}}}_{F^2 Q'^2} \\
& +\ldots~.
\end{split}
$}
\ee
We see that the only relevant operators are $FQ'$, $S^3$, $SFF$ and $FFF$.

Let us now deform the fixed point of \eref{dualC2A2b} by adding the singlets $M_{LL}$ and $F_L$ and turning on superpotential \eref{supdualC2A2a}.  Note that $M_{DD}$, $M_{UU}$, $*A$ and $Q$ are parts of the field $F$.   From the index \eref{indexSU9}, the terms in the second line obviously correspond to irrelevant operators.  Since $M_{LL}$ is a singlet that is added to the fixed point of \eref{dualC2A2b}, we have $R[M_{LL}] = \frac{2}{3}$, and so each term in the first line of \eref{supdualC2A2a} corresponds to an irrelevant operator; for example, $R[M_{LL}] +R[S^3] \approx \frac{2}{3} + \frac{179}{100} >2$.  The last term in \eref{supdualC2A2a} gives mass to the singlet $M_{LL}$ via the vacuum expectation value of $F_L$.  In summary, adding the singlets and turning on the deformation \eref{supdualC2A2a} makes the theory flow back to the original fixed point of \eref{dualC2A2b}.

In conclusion, theory \eref{C2A2} with superpotential \eref{supC2A2} and the dual theory \eref{dualC2A2a} with superpotential \eref{supdualC2A2a} flow to the same fixed point as that of theory \eref{dualC2A2b}.   As a result, the flavour symmetry of each of these theories is $SU(9) \times SU(2) \times U(1)^2$.  We emphasise again that, for theories \eref{C2A2} and \eref{dualC2A2a}, the $SU(9)$ global symmetry is not visible in the UV but is emergent in the IR.

\section{Conclusion and perspectives} \label{sec:conclusion}
A number of 4d $\CN=1$ gauge theories with interesting IR properties, such as flavour symmetry and supersymmetry enhancement, are proposed and studied.  The main approach that is used to construct such theories is to start with 4d $\CN=1$ gauge theories obtained by the compactification of 6d SCFTs on a torus with fluxes.  We then modify such theories by reducing the number of flavours as well as dropping or adding superpotential terms and flipping fields.  Although such a procedure leads to a number of interesting theories, supersymmetry or flavour symmetry enhancement is not guaranteed in the IR limit.  It would be nice to have a systematic method to produce such models.

Another interesting direction is to further study models similar to \eref{modelwEUSp}, namely those containing $E[USp(2N)]$ as a component, as well as its compactification on a circle to a 3d $\CN=2$ gauge theory with an appropriate monopole superpotential turned on.  As we pointed out in footnote \ref{Sfoldfootnote}, the construction of \eref{modelwEUSp} is in the same spirit of that of the 3d S-fold SCFTs \cite{Terashima:2011qi, Gang:2015wya, Assel:2018vtq, Garozzo:2018kra, Garozzo:2019hbf, Garozzo:2019ejm}, which possess 3d $\CN=3$ or $\CN=4$ supersymmetry. The dimensional reduction of $E[USp(2N)]$, as showed in \cite{Pasquetti:2019hxf}, has indeed a limit to the $T[SU(N)]$ theory used in the S-fold construction\footnote{This limit consists of two consecutive real mass deformations of the dimensional reduction of $E[USp(2N)]$. After the first deformation, we reach an intermediate theory called $M[SU(N)]$ which was introduced in \cite{Pasquetti:2019tix} exploiting a correspondence between the $\mathbb{S}^2\times \mathbb{S}^1$ partition function for 3d $\mathcal{N}=2$ theories and 2d CFT correlators in the free field realization \cite{Pasquetti:2019uop}. Also this $M[SU(N)]$ theory is suitable for being used as a building block to construct 3d $\mathcal{N}=2$ that generalise the S-fold models.}. Hence, the compactification of the 4d $\CN=1$ theories containing the $E[USp(2N)]$ building blocks on a circle would naturally give rise to the 3d $\CN=2$ analog of the aforementioned 3d $S$-fold SCFTs\footnote{Some constructions similar to the S-fold models but with a lower amount of supersymmetry have been studied in \cite{Garozzo:2019xzi}, where the building block used is a $U(N)$ gauge theory with $2N$ fundamental flavors and a monopole superpotential that lives on the duality domain wall of the 4d $\mathcal{N}=2$ $SU(N)$ gauge theory with $2N$ flavors \cite{Benini:2017dud,Floch:2015hwo}.}.  Recently there have been a proposal regarding a class of $\CN = 2$ $S$-fold solutions in Type IIB supergravity of the form $AdS_4 \times S^1 \times S^5$ involving S-duality twists of hyperbolic type along $S^1$ \cite{Guarino:2020gfe}.  It would be interesting to see if there is any connection between such a  3d $\CN=2$ analog in the large $N$ limit to this type of supergravity solutions.

\acknowledgments
We are indebted to Antonio Amariti, Sara Pasquetti and Alberto Zaffaroni for valuable discussions.  I.G. is partially supported by the INFN.  M.S. is partially supported by the ERC-STG grant 637844-HBQFTNCER, by the University of Milano-Bicocca grant 2016-ATESP0586, by the MIUR-PRIN contract 2017CC72MK003, and by the INFN.  G.Z. is partially supported by the ERC-STG grant 637844-HBQFTNCER and by the INFN.

\appendix

\section{Review of the $E[USp(2N)]$ theory}
\label{euspapp}
In this appendix, we review some properties of the $E[USp(2N)]$ theory, which was first introduced in \cite{Pasquetti:2019hxf} and further studied in \cite{Hwang:2020wpd}. The $E[USp(2N)]$ theory is a 4d $\mathcal{N}=1$ superconformal field theory that is realised as the IR fixed point of the following quiver theory\footnote{In comparison with figure 3 of \cite{Pasquetti:2019hxf}, the quiver for $E[USp(2N)]$ in this paper does not have the flipping fields for $D^{(N)}D^{(N)}$, and does not have an antisymmetric chiral multiplet under the rightmost square node $USp(2N)$.}:
\be
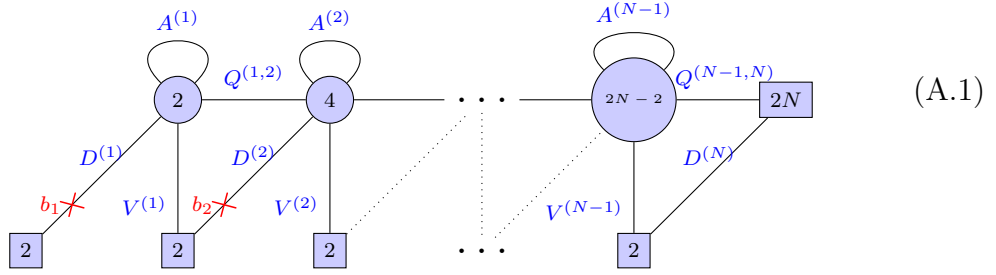
\label{euspfields} 
\begin{tikzpicture}[baseline]
\tikzstyle{every node}=[font=\scriptsize]
\node[draw, circle, fill=blue!20] (c2) at (0,0) {$2$};
\node[draw, circle, fill=blue!20] (c4) at (2,0) {$4$};
\node[draw=none] (tdots) at (4,0) {\Large $\ldots$};
\node[draw, circle, fill=blue!20] (c2nm2) at (6,0) {\fontsize{4pt}{4pt}\selectfont $2N-2$};
\node[draw, rectangle, fill=blue!20] (s2n) at (8,0) {$2N$};
\node[draw, rectangle, fill=blue!20] (s0) at (-2,-2) {$2$};
\node[draw,rectangle, fill=blue!20] (s2) at (0,-2) {$2$};
\node[draw, rectangle, fill=blue!20] (s4) at (2,-2) {$2$};
\node[draw=none] (bdots) at (4,-2) {\Large $\ldots$};
\node[draw, rectangle, fill=blue!20] (s2nm2) at (6,-2) {\selectfont $2$};
\draw[-] (s0) to node[above]{\blue $D^{(1)}$} node[near start, rotate=30]{\large\red $\times$}  node[near start, left]{\red $b_1$}  (c2);
\draw[-] (s2) to node[above]{\blue $D^{(2)}$} node[near start, rotate=30]{\large\red $\times$}  node[near start, left]{\red $b_2$} (c4);
\draw[-,dotted] (s4) to  (tdots);
\draw[-,dotted] (bdots) to  (c2nm2);
\draw[-] (s2nm2) to node[above]{\blue $D^{(N)}$}  (s2n);
\draw[-] (c2) to node[above]{\blue $Q^{(1,2)}$} (c4);
\draw[-] (c4) to  (tdots);
\draw[-] (tdots) to  (c2nm2);
\draw[-] (c2nm2) to node[above, xshift=0.1cm]{\blue $Q^{(N-1,N)}$} (s2n);
\draw[-] (c2) to node[left, near end]{\blue $V^{(1)}$} (s2);
\draw[-] (c4) to node[left, near end]{\blue $V^{(2)}$} (s4);
\draw[-,dotted] (tdots) to (bdots);
\draw[-] (c2nm2) to node[left, near end]{\blue $V^{(N-1)}$} (s2nm2);
\draw[black,solid] (c2) edge [out=45,in=135,loop,looseness=6] node[above]{\blue $A^{(1)}$} (c2);
\draw[black,solid] (c4) edge [out=45,in=135,loop,looseness=6] node[above]{\blue $A^{(2)}$} (c4);
\draw[black,solid] (c2nm2) edge [out=45,in=135,loop,looseness=3] node[above]{\blue $A^{(N-1)}$} (c2nm2);
\end{tikzpicture}
\ee
where each blue node labelled by an {\it even number} $m$ denotes the group $USp(m)$.  Here $D^{(n)}$ stand for the chiral multiplets represented by diagonal lines, $V^{(n)}$ stand for the chiral multiplets represented by vertical lines, and $A^{(n)}$ are the chiral multiplets in the rank-two antisymmetric representation $[0,1,0,\ldots,0]+[0,\ldots,0]= {\bf n(2n-1)}$ of $USp(2n)$.
The superpotential is taken to be
\begin{eqnarray}
W_{\eref{euspfields}}&=&\sum_{n=1}^{N-1}A^{(n)}\left(Q^{(n,n+1)}Q^{(n,n+1)}-Q^{(n-1,n)}Q^{(n-1,n)}\right)+\nn\\
&+&\sum_{n=1}^{N-1}V^{(n)}Q^{(n,n+1)}D^{(n+1)}+\sum_{n=1}^{N-1} b_n D^{(n)}D^{(n)}\, ,
\label{superpoteusp}
\end{eqnarray}
where we omitted contractions of indices, which are always performed using the antisymmetric tensor $J=\mathbb{I}_n\otimes i\,\gs_2$.

The manifest non-anomalous global symmetry of \eref{euspfields} is\footnote{It is worth noting that the $U(1)_\tau$ in this paper was referred to as $U(1)_t$ in the original reference \cite{Pasquetti:2019hxf}.  The reason that we change the notation $t$ to $\tau$ in this paper is to avoid the confusion with the fugacity $t$ in the index.}
\be
USp(2N)_x\times SU(2)^N\times U(1)_\fug\times U(1)_c\, .
\ee
This symmetry gets enhanced in the IR to
\be
USp(2N)_x\times USp(2N)_y\times U(1)_\fug\times U(1)_c\, ,
\label{euspsymmetry}
\ee
which is the non $R$-global symmetry of the $E[USp(2N)]$ theory.
The enhancement was argued in \cite{Pasquetti:2019hxf} by showing that the gauge invariant operators rearrange into representations of the enhanced $USp(2N)_y$ symmetry (\eg~ using the supersymmetric index) and by means of a duality, called of \emph{mirror-type} in \cite{Hwang:2020wpd}, which allows us to find a dual frame where $USp(2N)_y$ is manifest while $USp(2N)_x$ is emergent in the IR.


We schematically summarise the charges under the abelian symmetries of the chiral fields and a possible choice for the trial R-charge below:
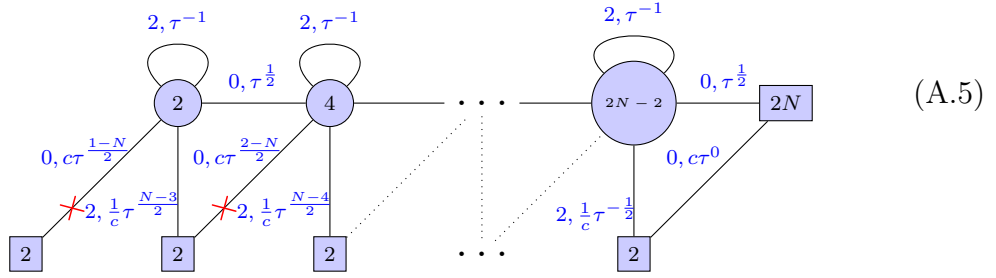
\be\label{euspfugacities} 
\begin{tikzpicture}[baseline]
\tikzstyle{every node}=[font=\scriptsize]
\node[draw, circle, fill=blue!20] (c2) at (0,0) {$2$};
\node[draw, circle, fill=blue!20] (c4) at (2,0) {$4$};
\node[draw=none] (tdots) at (4,0) {\Large $\ldots$};
\node[draw, circle, fill=blue!20] (c2nm2) at (6,0) {\fontsize{4pt}{4pt}\selectfont $2N-2$};
\node[draw, rectangle, fill=blue!20] (s2n) at (8,0) {$2N$};
\node[draw, rectangle, fill=blue!20] (s0) at (-2,-2) {$2$};
\node[draw,rectangle, fill=blue!20] (s2) at (0,-2) {$2$};
\node[draw, rectangle, fill=blue!20] (s4) at (2,-2) {$2$};
\node[draw=none] (bdots) at (4,-2) {\Large $\ldots$};
\node[draw, rectangle, fill=blue!20] (s2nm2) at (6,-2) {\selectfont $2$};
\draw[-] (s0) to node[above, xshift=-0.2cm]{\blue $0, c \tau^{\frac{1-N}{2}}$} node[near start, rotate=30]{\large\red $\times$}    (c2);
\draw[-] (s2) to node[above, xshift=-0.2cm]{\blue $0, c \tau^{\frac{2-N}{2}}$} node[near start, rotate=30]{\large\red $\times$}   (c4);
\draw[-,dotted] (s4) to  (tdots);
\draw[-,dotted] (bdots) to  (c2nm2);
\draw[-] (s2nm2) to node[above, xshift=-0.2cm]{\blue $0, c \tau^0$}  (s2n);
\draw[-] (c2) to node[above]{\blue $0, \tau^{\frac{1}{2}}$} (c4);
\draw[-] (c4) to  (tdots);
\draw[-] (tdots) to  (c2nm2);
\draw[-] (c2nm2) to node[above, xshift=0.1cm]{\blue $0, \tau^{\frac{1}{2}}$} (s2n);
\draw[-] (c2) to node[left, near end, xshift=0.2cm]{\blue $2, \frac{1}{c} \tau^{\frac{N-3}{2}}$} (s2);
\draw[-] (c4) to node[left, near end, xshift=0.2cm]{\blue $2, \frac{1}{c} \tau^{\frac{N-4}{2}}$} (s4);
\draw[-,dotted] (tdots) to (bdots);
\draw[-] (c2nm2) to node[left, near end, xshift=0.2cm]{\blue $2, \frac{1}{c} \tau^{-\frac{1}{2}}$} (s2nm2);
\draw[black,solid] (c2) edge [out=45,in=135,loop,looseness=6] node[above]{\blue $2, \tau^{-1}$} (c2);
\draw[black,solid] (c4) edge [out=45,in=135,loop,looseness=6] node[above]{\blue $2, \tau^{-1}$} (c4);
\draw[black,solid] (c2nm2) edge [out=45,in=135,loop,looseness=3] node[above]{\blue $2, \tau^{-1}$} (c2nm2);
\end{tikzpicture}
\ee
where each number before the comma (,) denotes the trial $R$-charge $U(1)_{R_0}$, and the powers of $c$ and $\tau$ denote the charges under $U(1)_c$ and $U(1)_\tau$ respectively. 

The operators of $E[USp(2N)]$ that are important for us are the following:
\begin{itemize}
\item the meson matrix $\mathsf{H}=Q^{(N-1,N)}Q^{(N-1,N)}$ transforming in the traceless antisymmetric representation $[0,1,0,\ldots, 0] = {\bf N(2N-1)-1}$ of $USp(2N)_x$;
\item the matrix $\mathsf{C}$ transforming in the traceless antisymmetric representation of the enhanced $USp(2N)_y$.  From the description \eref{euspfields} , where only $SU(2)^N \subset USp(2N)_y$ is manifest, $\mathsf{C}$ can be constructed by using the following branching rule of $USp(2N)_y \rightarrow SU(2)^N$:
\bes{
&{\bf N(2N-1)-1} \\
&\qquad \rightarrow (N-1)\times({\bf 1},\cdots,{\bf 1})\oplus \left[ ({\bf 2},{\bf 2},{\bf 1},\cdots,{\bf 1})\text{ }\oplus\text{ permutations} \right]\, ,
}
where $(N-1)\times({\bf 1},\cdots,{\bf 1})$ corresponds to the traces $J_{ab} A_{ab}^{(n)}$ with $n=1,\cdots,N-1$ and $a, b=1,\ldots, 2n$ are the indices of the fundamental representation $USp(2n)$, while $\left[ ({\bf 2},{\bf 2},{\bf 1},\cdots,{\bf 1})\text{ }\oplus\text{ permutations} \right]$ corresponds to the operators of the form 
\be
D^{(i)}\left(\prod_{l=i}^{j-1}Q^{(l,l+1)}\right)V^{(j)} 
\ee
with $i=1,\cdots,N-1$ and $j=i+1,\cdots,N-1$;
\item the operator $\Pi$ in the bifundamental representation of $USp(2N)_x\times USp(2N)_y$, which is constructed collecting operators charged under the manifest $SU(2)^N\subset USp(2N)_y$ in \eref{euspfields} according to the branching rule
\be
{\bf 2N}\rightarrow({\bf 2},{\bf 1},\cdots,{\bf 1})\oplus({\bf 1},{\bf 2},{\bf 1},\cdots,{\bf 1})\oplus\cdots\oplus({\bf 1},\cdots,{\bf 1},{\bf 2})\, ,
\ee
where these $N$ $SU(2)$ fundamentals are operators of the form $D^{(i)}\prod_{l=i}^{N-1}Q^{(l,l+1)}$ with $i=1,\cdots,N$;
\item the gauge invariant combinations constructed using the last vertical chiral multiplets, dressed with powers of the last antisymmetric chiral multiplets:
\be
\mathsf{M}_n=\left(A^{(N-1)}\right)^{n-1}V^{(N-1)}V^{(N-1)},\qquad n=1,\cdots,N-1\,,
\ee
which are singlets under the non-abelian symmetries;
\item the flipping fields $b_n$ for $n=1,\cdots,N-1$ which are also singlets under the non-abelian symmetries.
\end{itemize}
The transformation rules of these operators under the enhanced global symmetry \eqref{superpoteusp} are listed below:
\be \label{tablechargeEUSp}
\scalebox{1}{
\begin{tabular}{c|cccc|c}
{} & $USp(2N)_x$ & $USp(2N)_y$ & $U(1)_\fug$ & $U(1)_c$ & $U(1)_{R_0}$ \\ \hline
$\mathsf{H}$ & ${\bf N(2N-1)-1}$ & $\mathbf{1}$ & $1$ & 0 & 0 \\
$\mathsf{C}$ & $\mathbf{1}$ & ${\bf N(2N-1)-1}$ & $-1$ & 0 & 2 \\
$\Pi$ & $\bf N$ & $\bf N$ & 0 & $+1$ & 0 \\
$\mathsf{M}_n$ & $\mathbf{1}$ & $\mathbf{1}$ & $-n$ & $-2$ & $2(n+1)$ \\
$b_n$ & $\mathbf{1}$ & $\mathbf{1}$ & $N-n$ & $-2$ & 2
\end{tabular}}
\ee

$E[USp(2N)]$ is self-dual with a non-trivial map of the gauge invariant operators. More precisely, the duality interchanges the $USp(2N)_x$ and $USp(2N)_y$ symmetries and redefines the $U(1)_\fug$ symmetry and the trial R-symmetry, while it leaves $U(1)_c$ unchanged. Denoting with $R_\fug$ the mixing coefficient of $U(1)_\fug$ with $U(1)_{R_0}$, the action of the duality on these symmetries can be encoded in
\begin{eqnarray}
R_\tau \quad\leftrightarrow\quad 2-R_\tau\,.
\label{Rtmirror}
\end{eqnarray}
The operators are accordingly mapped as
\begin{eqnarray}
\mathsf{H}\quad&\leftrightarrow&\quad \mathsf{C}^\vee\nn\\
\mathsf{C}\quad&\leftrightarrow&\quad \mathsf{H}^\vee\nn\\
\Pi\quad&\leftrightarrow&\quad \Pi^\vee\nn\\
b_n\quad&\leftrightarrow&\quad \mathsf{M}_{N-n}^\vee\nn\\
\mathsf{M}_n\quad&\leftrightarrow&\quad b_{N-n}^\vee\, ,
\end{eqnarray}
where the superscript ${}^\vee$ labels the operators in the dual frame.

In the main text, we use the superconformal field theory $E[USp(2N)]$ as a building block to construct a more complicated model by gauging the $USp(2N)_x$ and $USp(2N)_y$ symmetries and coupling them to some additional matter fields.  For this purpose, it is useful to represent $E[USp(2N)]$ by a diagram where we explicitly show both its $USp(2N)$ global symmetries:
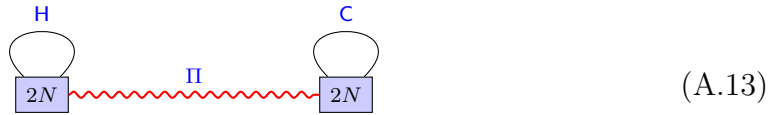
\be \label{blockEUSp}
\begin{tikzpicture}[baseline]
\tikzstyle{every node}=[font=\scriptsize]
\node[draw, rectangle, fill=blue!20] (cL) at (-2,0) {$2N$};
\node[draw, rectangle, fill=blue!20] (cR) at (2,0) {$2N$};
\draw[-, red, thick, snake it] (cL) to node[above]{\blue $\Pi$} (cR);
\draw[black,solid] (cL) edge [out=45,in=135,loop,looseness=6] node[above]{\blue $\mathsf{H}$} (cL);
\draw[black,solid] (cR) edge [out=45,in=135,loop,looseness=6] node[above]{\blue $\mathsf{C}$} (cR);
\end{tikzpicture}
\ee
where the left and right nodes correspond to $USp(2N)_x$ and $USp(2N)_y$ respectively.  We display explicitly the operator $\Pi$, $\mathsf{H}$ and $\mathsf{C}$.  We emphasise that these are composites of chiral fields in the quiver description \eref{euspfields}.  The other operators, such as $\mathsf{M}_n$ and $b_n$, which transform trivially under each $USp(2N)$ symmetry are omitted from the diagram.  

One important ingredient to analyse various models in the main text is the contribution of the $E[USp(2N)]$ block to the $\tr U(1)_R$ anomaly of each $USp(2N)$ gauge node. When the node that we are gauging corresponds to the manifest $USp(2N)_x$ symmetry, its contribution to $\tr U(1)_R$ is
\be
USp(2N)_x:\qquad\Tr\,U(1)_R\supset (N-1)\left(\frac{R_\tau}{2}-1\right)+(R_c-1)\, ,
\ee
where the first term is the contribution of $Q^{(N-1,N)}$, while the second term is the contribution of $D^{(N)}$. On the other hand, for the $USp(2N)_y$ symmetry, it is not convenient to use the quiver description \eref{euspfields} of $E[USp(2N)]$, since $USp(2N)_y$ is not manifest in that description but is emergent in the IR. Nevertheless, we can take advantage of the self-duality of $E[USp(2N)]$.  Specifically, we can compute the contribution to the $U(1)_R$ anomaly of the gauged $USp(2N)_y$ node using its dual frame where this symmetry is manifest. Using \eqref{Rtmirror} we find that such a contribution is
\bes{
USp(2N)_y:\qquad\Tr\,U(1)_R &\supset  (N-1)\left(\frac{2-R_\tau}{2}-1\right)+(R_c-1) \\
& \quad = -\frac{N-1}{2}R_\fug+(R_c-1)\, .
}

Another important result that we used in the main text is that the contribution to the $U(1)_\tau^3$ 't Hooft anomaly of the $E[USp(2N)]$ block is zero for any $N$:
\begin{eqnarray}
\Tr\,U(1)_\fug^3&\supset& \sum_{n=1}^{N-1}n(2n-1)(-1)^3+\sum_{n=1}^{N-1}4n(n+1)\left(\frac{1}{2}\right)^3+\sum_{n=1}^N4n\left(\frac{n-N}{2}\right)^3+\nn\\
&+&\sum_{n=1}^{N-1}4n\left(\frac{N-n-2}{2}\right)^3+\sum_{n=1}^{N-1}\left(-\frac{n-N}{2}\right)^3=0\,,
\label{u1tcube}
\end{eqnarray}
where in order we have written the contributions of $A^{(n)}$, $Q^{(n,n+1)}$, $D^{(n)}$, $V^{(n)}$ and $b_n$.

\section{Supersymmetric index}\label{app:index}
In this appendix we briefly summarise basic notion of the supersymmetric index for 4d $\mathcal{N}=1$ SCFTs \cite{Kinney:2005ej,Romelsberger:2005eg,Dolan:2008qi}; see also \cite{Rastelli:2016tbz} for a more comprehensive review.  We follow closely the exposition of the latter reference.

The index of a 4d $\CN=1$ SCFT is a refined Witten index of the theory quantized on $\S^3\times {\mathbb R}$,
\be
\mathcal{I}=\Tr(-1)^F e^{-\beta \delta} e^{-\mu_i \mathcal{M}_i}, \qquad  \delta=\half \left\{\mathcal{Q},\mathcal{Q}^{\dagger}\right\}~,
\ee
where $\mathcal{Q}$ is one of the Poincar\'e supercharges; $\mathcal{Q}^{\dagger}=\mathcal{S}$ is the conjugate conformal supercharge; $\mathcal{M}_i$ are $\mathcal{Q}$-closed conserved charges, and $\mu_i$ are their chemical potentials. All the states contributing to the index with non-vanishing weight have $\delta=0$; this renders the index independent of $\beta$.

For $\mathcal{N}=1$ SCFTs, the supercharges are 
\bes{
\left\{\mathcal{Q}_{\alpha},\,\mathcal{S}^{\alpha} = \mathcal{Q}^{\dagger\alpha}\widetilde{\mathcal{Q}}_{\dot{\alpha}},\,\widetilde{\mathcal{S}}^{\dot{\alpha}} = \widetilde{\mathcal{Q}}^{\dagger\dot{\alpha}}\right\}~,
} 
where $\alpha=\pm$ and $\dot{\alpha}=\dot{\pm}$ are respectively the $SU(2)_1$ and $SU(2)_2$ indices of the isometry group $Spin(4)=SU(2)_1 \times SU(2)_2$ of $S^3$.
For definiteness, let us choose $\mathcal{Q}=\widetilde{\mathcal{Q}}_{\dot{-}}$. With this particular choice, it is common to define the index as a function of the following fugacities
\be
\mathcal{I}\left(p,q\right)=\Tr(-1)^F p^{j_1 + j_2 +\half r} q^{j_2 - j_1 +\half r}.
\ee
where $p$ and $q$ are fugacities associated with the supersymmetry preserving squashing of the $S^3$ \cite{Dolan:2008qi}. Indeed, even if the dimension of the bosonic part of the 4d $\mathcal{N}=1$ superconformal algebra is four, the number of independent fugacities that we can turn on in the index is two because of the constraints $\gd=0$ and $[\mathcal{M}_i,\mathcal{Q}]=0$. A possible choice for the combinations of the bosonic generators that satisfy these requirements is $\pm j_1+j_2+\frac{r}{2}$, where $j_1$ and $j_2$ are the Cartan generators of $SU(2)_1$ and $SU(2)_2$, and $r$ is the generator of the $U(1)_r$ $R$-symmetry.  In the main text, we write
\be
t= (p q)^{\frac{1}{2}}~, \qquad y = \left( \frac{p}{q} \right)^{\frac{1}{2}}~.
\ee

The index counts gauge invariant operators that can be constructed from modes of the fields. The latter are usually referred to as `letters' in the literature. The single-letter index for a vector multiplet and a chiral multiplet $\chi(\mathbf{R})$ transforming in the $\mathbf{R}$ representation of the gauge$\times$flavour group is
\bes{
i_V \left(t,y,U\right) & =  \frac{2t^2-t(y+y^{-1})}{(1-t y)(1-t y^{-1})} \chi_{\mathrm{adj}}\left(U\right), \nn\\
i_{\chi(\mathbf{R})}\left(t,y,U,V\right) & = 
\frac{t^{r} \chi_{\mathbf{R}} \left(U,V\right) - t^{2-r} \chi_{\bar{\mathbf{R}}} \left(U,V\right)}{(1-t y )(1-t y^{-1})},
}
where $\chi_{\mathbf{R}} \left(U,V\right)$ and $\chi_{\bar{\mathbf{R}}} \left(U,V\right)$ denote the characters of $\mathbf{R}$ and the conjugate representation $\bar{\mathbf{R}}$, with $U$ and $V$ gauge and flavour group matrices, respectively.

The index can then be obtained by symmetrising of all of such letters and then projecting them to gauge singlets by integrating over the Haar measure of the gauge group. This takes the general form
\be
\mathcal{I} \left(t,y,V\right)=\int \left[dU\right] \prod_{k} \PE\left[i_k\left(t,y,U,V\right)\right],
\ee
where $k$ labels the different multiplets in the theory, and $\PE[i_k]$ is the plethystic exponential of the single-letter index of the $k$-th multiplet, responsible for generating the symmetrisation of the letters. It is defined by
\be
\PE\left[i_k\left(t,y,U,V\right)\right] = \exp \left[  \sum_{n=1}^{\infty} \frac{1}{n} i_k\left(t^n,y^n,U^n,V^n\right) \right].
\ee

For definiteness, let us discuss a specific example of the $SU(N_c)$ gauge group.  The contribution of a chiral superfield in the fundamental representation $\mathbf{N_c}$ or antifundamental representation $\bar{\mathbf{N_c}}$ of $SU(N_c)$ with $R$-charge $r$ can be written in terms of elliptic gamma functions, as follows
\bes{
\PE\left[i_{\chi(\mathbf{N_c})} \left(t, y,U\right)\right] &= \prod_{i=1}^{N_c} \Gamma_e \left( t^r  z_i \right) ~, \quad 
\PE\left[i_{\chi(\bar{\mathbf{N_c}})}  \left(t, y,U\right)\right] = \prod_{i=1}^{N_c} \Gamma_e \left( t^r  z^{-1}_i \right)  \\
\Gamma_e(z)\equiv \Gamma\left(z;t,y\right) & = \prod_{n,m=0}^{\infty} \frac{1-(t y)^{n+1} (ty^{-1})^{m+1} z^{-1}}{1- (ty)^n  (t y^{-1})^m z}~,
}
where $\{z_i\}$, with $i=1,...,N_c$ and $\prod_{i=1}^{N_c} z_i=1$, are the fugacities parameterising the Cartan subalgebra of $SU(N_c)$.  We will also use the shorthand notation
\be
\Gamma_e \left(u z^{\pm n} \right)=\Gamma_e \left(u z^{n} \right)\Gamma_e \left(u z^{-n} \right).
\ee
On the other hand, the contribution of the vector multiplet in the adjoint representation of $SU(N_c)$, together with the $SU(N_c)$ Haar measure, is
\be
\frac{\kappa^{N_c-1}}{N_c!} \oint_{\mathbb{T}^{N_c}} \prod_{i=1}^{N_c-1} \frac{dz_i}{2\pi i z_i} \prod^{N_c}_{k\neq \ell} \frac1{\Gamma_e(z_k z_\ell^{-1})} \cdots~,
\ee
where the dots denote that it will be used in addition to the full matter multiplets transforming in representations of the gauge group. The integration contour is taken over the maximal torus of the gauge group and $\kappa$ is the index of $U(1)$ free vector multiplet defined as
\be
\kappa = (t y; ty)( t y^{-1} ; t y^{-1}),
\ee
with $(a;b) = \prod_{n=0}^\infty \left( 1-ab^n \right)$ the $q$-Pochhammer symbol.  A similar discussion for the $USp(2N_c)$ gauge group can be found in appendix B of \cite{Pasquetti:2019hxf}.

At the superconformal fixed point, we employ the superconformal symmetry to extract the information about the states. Although the index counts states up to cancellations due to recombinations of various short superconformal multiplets to long multiplets, it has been shown in \cite{Beem:2012yn} that at low orders of the expansion in $t$ the index reliably contains information about certain important operators.  In particular, at order $t^2=pq$, one obtains the difference between the marginal operators and the conserved currents.  We extensively utilise the result of the computation at this order in the main text.

\subsection{Example: Index of theory \eref{model12meq4}} \label{sec:model12meq4}
Let us now discuss how to obtain the index of theory \eref{model12meq4} with the charge assignment given in \eref{chargemodel1}.  A technical problem here is that the chiral field $D$ carries zero $R$-charge, which causes the problem in obtaining the power series in $t$ of the index \eref{indexmodel1}.  One way to circumvent this problem is to assign an extra $U(1)$ symmetry, which we shall refer to as a `fake' symmetry and is denoted by $U(1)_f$, so that it mixes with the $R$-symmetry in such a way that the chiral fields originally carrying zero $R$-charge now have positive $R$-charge.  As an example, we can assign the $U(1)_f$ charge of $D$ to be $2$ such that the new $R$-charge of $D$ is $t^{\frac{8}{9}}$. 
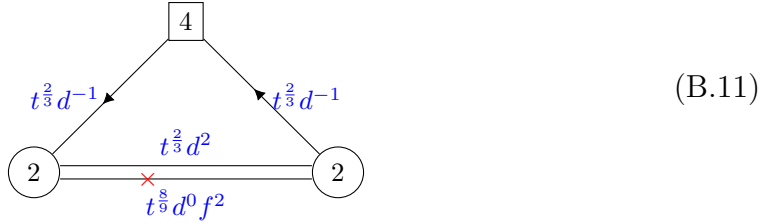
\be \label{fakeU1}
\begin{tikzpicture}[baseline]
\tikzstyle{every node}=[font=\footnotesize]
\node[draw, circle] (node1) at (-2,-1) {$2$};
\node[draw, circle] (node2) at (2,-1) {$2$};
\node[draw, rectangle] (sqnode) at (0,1) {$4$};
 \draw[transform canvas={yshift=-2.5pt}] (node1) to  node[below] {\blue $t^{\frac{8}{9}} d^0 f^2$} node[below, xshift=-0.5cm] {} node[xshift=-0.5cm] {$ \red \times$} (node2) ;
  \draw[transform canvas={yshift=2.5pt}] (node1) to node[above] {\blue $t^{\frac{2}{3}} d^2$} (node2);
\draw[draw=black,solid, -<-=0.5]  (node1) to node[left] {\blue $t^{\frac{2}{3}} d^{-1}$} (sqnode);
\draw[draw=black,solid, -<-=0.5]  (sqnode) to node[right] {\blue $t^{\frac{2}{3}} d^{-1}$} (node2);
\end{tikzpicture}
\ee
Of course, the flipping field $F_D$ now has $U(1)_f$ charge $-4$ and $R$-charge $2- \frac{16}{9} = \frac{2}{9}$.  The main idea is to compute the index using the charge assignment \eref{fakeU1} as a power series in $t$, and then set $f= t^{-\frac{4}{9}}$ to obtain the index of the original theory \eref{chargemodel1}.

The index of \eref{fakeU1} can be written as
\bes{
\CI_{\eref{fakeU1}} (t, y; \vec u, f) &= \frac{\kappa^{2}}{2!2!} \oint_{|v|=1}  \frac{dv}{2\pi i v}  \frac{1}{\Gamma_e(v^2)} \oint_{|w|=1}  \frac{dw}{2\pi i w}  \frac{1}{\Gamma_e(w^2)} \times \\
& \CI_U(v,w,d)  \CI_D(v,w,f) \CI_{L}(v, \vec u, d) \CI_{R} (w, \vec u, d) \CI_{F_D} (f) ~,
}
where we have suppressed the variables $t$ and $y$ in the argument of each contribution on the right hand side, and
\bes{
\CI_U(v,w,d) &= \prod_{i, j =\pm 1} \Gamma_e (t^\frac{2}{3} v^i w^j d^2) ~, \\
\CI_D(v,w,f) &=   \prod_{i, j =\pm 1} \Gamma_e (t^\frac{8}{9} v^i w^j f^2) ~, \\
\CI_L(v,\vec u,d) &=   \prod_{i = \pm 1} \prod_{\alpha=1,\ldots 4} \Gamma_e (t^\frac{2}{3} v^i u^\alpha d^{-1})~,  \\
\CI_R(w,\vec u,d) &=   \prod_{i = \pm 1} \prod_{\alpha=1,\ldots 4} \Gamma_e (t^\frac{2}{3} w^i u^{-\alpha} d^{-1})~, \\
\CI_{F_D}(f) &=  \Gamma_e (t^{\frac{2}{9}} f^{-4})~.
}
The expression $\CI_{\eref{fakeU1}} (t, y; \vec u, f)$  can then be computed as a power series of $t$.  The index of theory \eref{model12meq4} with the charge assignment given in \eref{chargemodel1} is therefore
\be
\CI_{\eref{model12meq4}} (t, y; \vec u, f=t^{-\frac{4}{9}}) = \Gamma_e (t^{\frac{2}{3}} d^2) \times \CI_{\eref{indexmodel1}} (t, y; \vec u, d)~, 
\ee
where $\CI_{\eref{indexmodel1}}(t, y; \vec u, d)$ is the index given by \eref{indexmodel1} and the first factor is the contribution from the free chiral field corresponding to the operator $UD$.  

Alternatively, we can also flip the operator $UD$ by introducing the flipping field $F_{UD}$ with superpotential term $F_{UD} UD$.  The contribution of $F_{UD}$ to the index is
\be
\CI_{F_{UD}}(d,f) = \Gamma_e (t^{\frac{4}{9}} d^{-2} f^{-2})~.
\ee
The index $\CI_{\eref{indexmodel1}}(t, y; \vec u, d)$ can then be obtained by first computing a power series in $t$ of the following expression:
\be
\CI_{F_{UD}}(d,f) \CI_{\eref{fakeU1}} (t, y; \vec u, f)
\ee
and then set $f=t^{-\frac{4}{9}}$.

\bibliographystyle{ytphys}
\bibliography{ref}
\end{document}